\documentclass[aps,preprint,superscriptaddress]{revtex4}

\usepackage[english]{babel}
\usepackage{comment}
\usepackage{orcidlink}

\usepackage{amsmath}
\usepackage{graphicx}
\usepackage{hyperref}
\usepackage{aas_macros}
\usepackage[dvipsnames]{xcolor}
\hypersetup{colorlinks=true, citecolor=Plum, linkcolor=Plum,
urlcolor=Plum}

\begin{document}

\title{deci-Hz Gravitational Wave Observations\\on the Moon and Beyond}

\author{Emanuele~Berti\,\orcidlink{0000-0003-0751-5130}}
\email{berti@jhu.edu}
\affiliation{William H. Miller III Department of Physics and Astronomy, Johns Hopkins University, Baltimore, Maryland 21218, USA}

\author{Marica~Branchesi\,\orcidlink{0000-0003-1643-0526}}
\affiliation{Gran Sasso Science Institute (GSSI), I-67100 L’Aquila, Italy}
\affiliation{INFN, Laboratori Nazionali del Gran Sasso, I-67100 Assergi, Italy}

\author{Alessandra~Buonanno\,\orcidlink{0000-0002-5433-1409}}
\affiliation{Max Planck Institute for Gravitational Physics (Albert Einstein Institute), D-14476 Potsdam, Germany}
\affiliation{Department of Physics, University of Maryland, College Park, MD 20742, USA}

\author{Alessandra~Corsi\,\orcidlink{0000-0001-8104-3536}}
\affiliation{William H. Miller III Department of Physics and Astronomy, Johns Hopkins University, Baltimore, Maryland 21218, USA}

\author{Daniel~J.~D'Orazio\,\orcidlink{0000-0002-1271-6247}}
\affiliation{Space Telescope Science Institute, 3700 San Martin Drive, Baltimore, MD 21218, USA 3}
\affiliation{William H. Miller III Department of Physics and Astronomy, Johns Hopkins University, Baltimore, Maryland 21218, USA}
\affiliation{Niels Bohr Institute, Blegdamsvej 17, 2100 Copenhagen, Denmark} 

\author{Jan~Harms\,\orcidlink{0000-0002-7332-9806}}
\affiliation{Gran Sasso Science Institute (GSSI), I-67100 L’Aquila, Italy}
\affiliation{INFN, Laboratori Nazionali del Gran Sasso, I-67100 Assergi, Italy}

\author{Jason~M.~Hogan\,\orcidlink{0000-0003-1218-2692}}
\affiliation{Department of Physics, Stanford University, Stanford, California 94305, USA}

\author{Francesco~Iacovelli\,\orcidlink{0000-0002-4875-5862}}
\affiliation{William H. Miller III Department of Physics and Astronomy, Johns Hopkins University, Baltimore, Maryland 21218, USA}

\author{Karan~Jani\,\orcidlink{0000-0003-1007-8912}}
\affiliation{Vanderbilt Lunar Labs Initiative, Vanderbilt University, Nashville, Tennessee,  37235, USA}
\affiliation{Department of Physics \& Astronomy, Vanderbilt University, Nashville, Tennessee,  37235, USA}

\author{Marc~Kamionkowski\,\orcidlink{0000-0001-7018-2055}}
\affiliation{William H. Miller III Department of Physics and Astronomy, Johns Hopkins University, Baltimore, Maryland 21218, USA}

\author{Kentaro~Komori\,\orcidlink{0000-0002-4092-9602}}
\affiliation{Research Center for the Early Universe (RESCEU), The University of Tokyo, 7-3-1 Hongo, Bunkyo-ku, Tokyo 113-0033, Japan}
\affiliation{Department of Physics, The University of Tokyo, 7-3-1 Hongo, Bunkyo-ku, Tokyo 113-0033, Japan}

\author{Konstantinos~Kritos\,\orcidlink{0000-0002-0212-3472}}
\affiliation{William H. Miller III Department of Physics and Astronomy, Johns Hopkins University, Baltimore, Maryland 21218, USA}

\author{Andrea~Maselli\,\orcidlink{0000-0001-8515-8525}}
\affiliation{Gran Sasso Science Institute (GSSI), I-67100 L’Aquila, Italy}
\affiliation{INFN, Laboratori Nazionali del Gran Sasso, I-67100 Assergi, Italy}

\author{M.~Coleman~Miller\,\orcidlink{0000-0002-2666-728X}}
\affiliation{Department of Astronomy and Joint Space-Science Institute, University of Maryland, College Park, MD 20742-2421 USA}

\author{Chiara~M.~F.~Mingarelli\,\orcidlink{0000-0002-4307-1322}}
\affiliation{Department of Physics, Yale University, New Haven, 06520, CT, USA}

\author{Volker~Quetschke\,\orcidlink{0000-0002-8012-4868}}
\affiliation{Department of Physics and Astronomy, University of Texas Rio Grande Valley, Brownsville, Texas 78520, USA}

\author{B.~S.~Sathyaprakash\,\orcidlink{0000-0003-3845-7586}}
\affiliation{Institute for Gravitation and the Cosmos, The Pennsylvania State University, University Park, PA 16802, USA}
\affiliation{Department of Physics, The Pennsylvania State University, University Park, PA 16802, USA}
\affiliation{Department of Astronomy \& Astrophysics, The Pennsylvania State University, University Park, PA 16802, USA}

\author{David~H.~Shoemaker\,\orcidlink{0000-0002-4147-2560}}
\affiliation{Kavli Institute, Massachusetts Institute of Technology, Cambridge, Massachusetts 02139 USA}

\author{Joseph~Silk\,\orcidlink{0000-0002-1566-8148}}
\affiliation{William H. Miller III Department of Physics and Astronomy, Johns Hopkins University, Baltimore, Maryland 21218, USA}
\affiliation{Institut d'Astrophysique de Paris, CNRS and Sorbonne University,  75014 Paris, France}

\author{Jacob~P.~Slutsky}
\affiliation{Gravitational Astrophysics Laboratory, NASA Goddard Space Flight Center, Greenbelt, Maryland 20771 USA}

\author{James~Ira~Thorpe\,\orcidlink{0000-0001-9276-4312}}
\affiliation{Gravitational Astrophysics Laboratory, NASA Goddard Space Flight Center, Greenbelt, Maryland 20771 USA}

\author{James~Trippe\,\orcidlink{0000-0003-1525-6300}}
\affiliation{Department of Electrical and Computer Engineering, Vanderbilt University, Nashville, Tennessee 37235, USA}

\author{Daniele~Vetrugno\,\orcidlink{0000-0003-0937-1468}}
\affiliation{Max Planck Institute for Gravitational Physics (Albert Einstein Institute), 30167 Hannover, Germany}
\affiliation{Leibniz University Hannover, 30167 Hannover, Germany}

\author{Stefano~Vitale\,\orcidlink{0000-0002-2427-8918}}
\affiliation{Department of Physics, University of Trento,  38123 Povo, Trento, Italy}

\begin{abstract}
  This document summarizes talks and discussions from the workshop  ``deci-Hz Gravitational Wave Observations on the Moon and Beyond'' that took place at Johns Hopkins University between September~1 and September~3, 2025. The workshop focused on experimental proposals to observe gravitational waves in the deci-Hz band, including lunar detectors, laser interferometers in space, and atom interferometry; gravitational wave sources in the deci-Hz frequency band; and the multi-messenger and multi-band astronomy that would be enabled by these observations.
\end{abstract}

\maketitle

\tableofcontents

\clearpage

\section{Introduction}

The international scientific community is embarked on the design of next-generation experiments to vastly expand the reach of current gravitational wave (GW) detectors. Major efforts to expand beyond the reach of the LIGO-Virgo-KAGRA network \cite{LIGOScientific:2014pky,VIRGO:2014yos,Aso:2013eba} on the ground include Cosmic Explorer (CE)~\cite{Reitze:2019iox, Evans:2021gyd} in the US and the Einstein Telescope (ET)~\cite{Punturo:2010zz} in Europe. 
The space-borne Laser Interferometer Space Antenna (LISA)~\cite{Colpi:2024hlh}, scheduled for launch in 2035,
will be sensitive to GWs in the milli-Hz. Pulsar Timing Arrays (PTAs) have recently provided evidence for a nanohertz GW background~\cite{NANOGrav:2023gor, EPTA:2023fyk, Reardon:2023gzh, Xu:2023wog, Miles:2024seg, NANOGrav:2024tnd}; the future sensitivity of these searches is growing as ever more radio telescopes of ever greater sensitivity are added.

The deci-Hz frequency band has a well-known abundance of astrophysical and cosmological GW sources (see e.g.~\cite{Mandel:2017pzd,Sedda:2019uro,2021hgwa.bookE..50I,Ajith:2024mie}), but building sensitive GW detectors in this frequency range presents formidable technical challenges. 

This document is a summary of the workshop ``deci-Hz Gravitational Wave Observations on the Moon and Beyond''~\cite{deciHz}, that took place at Johns Hopkins University between September 1 and September 3, 2025.
The goal of the workshop was to stimulate discussions between the experimental communities developing a variety of experimental avenues to deci-Hz GW detection and theorists working on the corresponding GW sources. 

The workshop brought together experts from three different experimental areas that could lead to sensitive detectors in the deci-Hz band, namely:

\begin{itemize}
\item[(1)] {\bf GW detectors on the Moon, or seismometers that would observe the excitation of the Moon's body resonances by GWs;}

\item[(2)] {\bf Space interferometer concepts similar to LISA;}

\item[(3)] {\bf Atom interferometry in space.}
\end{itemize}

The event had two main goals: a lively and constructive open discussion of the scientific advantages and technical challenges of each approach, and the discussion of realistic timelines for each of them. 

Here we report the presentations and discussions that occurred at the workshop, in chronological order, with only minor editing. We hope that the discussions reported in this document will prompt more conversations and follow-up workshops to develop complementary experimental approaches for deci-Hz GW astronomy. 

\begin{figure}[t]
    \centering
    \includegraphics[width=\linewidth]{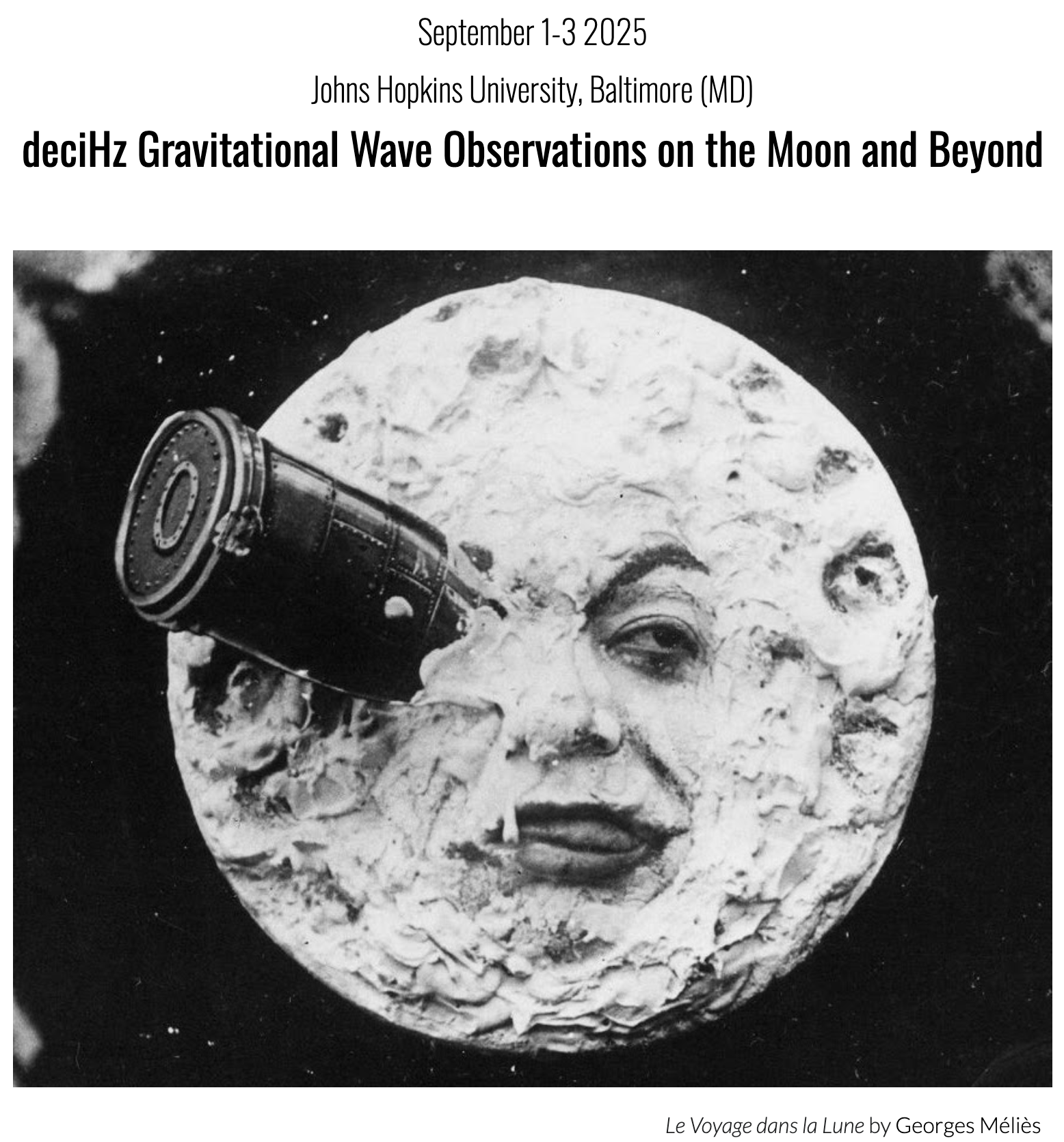}\\
    \caption{The workshop logo (a frame from the movie ``Le Voyage dans a Lune'' by George M\'eli\`es). See \cite{deciHz} for more information.}
    \label{fig:all_asds}
\end{figure}

\clearpage

\section{Monday morning -- Where deci-Hz fits in the future GW landscape}

\subsection{David Shoemaker: GWIC and the Landscape}

The Gravitational-Wave International Committee (GWIC~\cite{GWIC}) was established in 1997 to coordinate and support international efforts in GW science. It promotes collaboration among major projects, facilitates long-range planning, encourages the development of new technologies, and serves as a representative voice for the community. Membership includes ground-based interferometers such as LIGO, Virgo, KAGRA, NS-BH, and ET, as well as space missions like LISA, DECIGO, and TianQin, along with pulsar timing arrays and theory groups.
Over the years, GWIC has played an important role in organizing workshops, publishing roadmaps, and engaging with funding agencies. One of its major contributions is the 2021 Nature roadmap~\cite{Bailes:2021tot}, which outlined key scientific questions such as the formation and evolution of BHs, the physics of neutron stars (NSs), the role of dark matter and dark energy, and potential signatures of quantum gravity. Through the Gravitational-Wave Agencies Correspondents (GWAC), GWIC also maintains direct connections with major funding institutions including NASA, ESA, NSF, CNRS, ARC, and many others.

While GWIC has strongly advanced the development of kilometer-scale ground detectors, Pulsar Timing Arrays, and space-based interferometers, relatively little attention has been devoted to the deci-Hz frequency band. DECIGO~\cite{DECIGO} is already a GWIC member in an early stage of development, and TianQin~\cite{TianQin} was recently admitted in 2025, but the committee’s roadmaps have so far given only passing mention to this regime. An ad-hoc subcommittee has been formed to consider next generation space-based GW detectors, and the ESA Vision 2050 long-term planning activity~\cite{ESA_Voyage2050} is coordinated with that GWIC subcommittee. 

GWIC has not explored to date alternatives to laser interferometry, such as atom interferometry.

One possible outcome of this workshop would be a consensus on whether GWIC should take a more active role in supporting deci-Hz initiatives, whether a dedicated roadmap is needed, and how new projects in this band might engage with the committee.

A project interested in exploring membership in GWIC should contact the Chair or Executive Secretary~\cite{BailesShoemaker_contact}. GWIC meets yearly (at a minimum) and a discussion of possible membership can be arranged upon request. Projects should have recognition and some resources from appropriate funding agencies.    According to the Bylaws, a 2/3 vote of the current membership will admit a new project. 

\subsection{B.~S.~Sathyaprakash: Cosmic Explorer}

Cosmic Explorer (CE) is envisioned as the U.S. flagship next-generation GW observatory~\cite{Reitze:2019iox, Evans:2021gyd}. With planned baselines of 20\,km and 40\,km, CE aims to extend the reach of GW astronomy by more than an order of magnitude beyond Advanced LIGO. Its science goals span astrophysics, cosmology, nuclear physics, and tests of fundamental physics~\cite{Gupta:2023lga}. The publications  cited above give the current status of CE, its major science targets, and the opportunities and challenges that shape the project (see also~\cite{Evans:2023euw}). A sister concept for Europe, ET, is discussed below. We focus here on CE, but many of our comments are relevant also for ET. The network of CE, ET, and less sensitive instruments will revolutionize the search for GWs in the terrestrially accessible frequency range.

\noindent
\textbf{Status and timeline.}
The CE project is now at the conceptual development stage including site evaluation. In spring 2023 the National Science Foundation’s Mathematical and Physical Sciences Advisory Committee (MPSAC) established a subcommittee on next-generation GW facilities (ngGW)~\cite{Kalogera2024_MPSAC_ngGW_Report}. The CE project submitted a White Paper in response to the ngGW committee's call for letters of interest in pursuing a concept for an observatory that is ten times more sensitive than Advanced LIGO~\cite{Evans:2023euw}.  The committee concluded its study in the spring of 2024 and recommended CE considering two plausible scenarios~\cite{Kalogera2024_MPSAC_ngGW_Report}: if ET is constructed in Europe, the baseline would be a single 40\,km facility (CE40) in the United States; if ET is not realized, a network consisting of CE40 and a companion 20\,km facility (CE20) is envisaged. Furthermore, the committee emphasized the importance of successful construction and operation of the LIGO-India observatory for successful electromagnetic (EM) follow-up of GW events in the CE-ET epoch. 

CE will have a substantial geographical footprint due to its long, L-shaped arms planned to be on the Earth's surface. Site evaluation is currently underway, with both 40\,km and 20\,km options in play. The process includes extensive engagement with local communities and indigenous groups. The project considers ensuring responsible interaction with potential host communities a central priority, covering not only construction and commissioning but also eventual decommissioning and land return. Lessons from the Thirty Meter Telescope controversy emphasize the importance of consent and local support. Some communities have expressed enthusiasm, but the outcome will inevitably involve difficult trade-offs, as only one or two sites can ultimately be selected and the decision will be made by the funding agency.

\noindent
\textbf{Detector networks and localization.}
The configuration of CE relative to global facilities has strong implications for science. Exhaustive studies carried out for the CE White Paper and Horizon Study~\cite{Evans:2021gyd} show that while a single detector can localize long-duration coalescing-binary signals, three non-collocated sites are required for most sources~\cite{Borhanian:2022czq, Kalogera:2021bya}. In this context LIGO-India could provide powerful complementarity~\cite{Unnikrishnan:2013qwa, Pandey:2024mlo}: with two next-generation U.S. facilities plus LIGO-India, or CE40 and ET, multi-messenger follow-up would be possible up to a redshift of $z \sim 0.3$. However, pushing to higher redshifts ($z \sim 3$ for NSs and $z \sim 50$ for BHs) would require additional detectors. Underground facilities or space-based detectors could also mitigate limitations from seismically-induced gravitational acceleration fluctuations ~\cite{ET:2019dnz, Sathyaprakash:2012jk, Abac:2025saz, Colpi:2024hlh}.

\noindent
\textbf{Black hole and neutron star populations.}
CE will probe BH populations across cosmic history, addressing questions about the origin of the first BHs, the role of light seeds in the growth of supermassive BHs, and the connection between BH mergers and galaxy evolution. Similarly, CE will chart the neutron-star population, asking whether its properties at high redshift differ from those observed locally.

\noindent
\textbf{Nuclear equation of state.}
A central science goal is constraining the neutron-star equation of state (EOS). CE will measure the maximum mass ($M_{\rm max}$), minimum radius ($R_{\rm min}$), and mass-dependent tidal deformability~\cite{Huxford:2023qne, Kashyap:2025cpd, Khadkikar:2025ith}. Observations of post-merger oscillations may distinguish between nucleonic, hyperonic, and quark matter EOS models~\cite{Chatziioannou:2020pqz, Bauswein:2015vxa, Prakash:2023afe}, and may even reveal redshift evolution of EOS constraints. A 20\,km facility tuned with signal recycling could be optimized for sensitivity at a few kilohertz, enabling the detection of potential phase transitions in neutron-star cores~\cite{Evans:2021gyd}.

\noindent
\textbf{Multi-messenger astrophysics.}
CE will revolutionize multi-messenger astronomy (MMA). Joint GW and EM observations will clarify the progenitors of short gamma-ray bursts, determine whether all binary NS mergers produce kilonovae, and constrain the yields of $r$-process nucleosynthesis~\cite{Corsi:2024vvr}. MMA detections will also provide new measurements of the Hubble constant and other cosmological parameters, and will reveal the astrophysical environments of mergers~\cite{Branchesi:2023mws}. CE is engaging with the EM community to ensure coordinated observing strategies, recognizing the challenges of telescope time allocation and funding. Analogies to NASA’s \emph{Swift} satellite highlight the need for rapid EM response, with observations on minute timescales~\cite{Ronchini:2022gwk}.

\noindent
\textbf{Fundamental physics.}
CE offers opportunities for fundamental physics beyond the standard astrophysical program. Tests of general relativity, while not the primary focus, will be strengthened through precise measurements of inspiral and ringdown phases. Exotic physics targets include the influence of dark matter candidates (e.g., WIMPs or axion-like particles) on orbital dynamics, BH superradiance, and the possible transmutation of NSs into low-mass BHs.

\noindent
\textbf{Challenges and limitations.}
While CE and ET represent the pinnacle of terrestrial laser-interferometer GW technology, their reach is ultimately curtailed by their placement on the seismically noisy earth. A key obstacle for terrestrial observatories is Newtonian noise due to gravitational acceleration fluctuations, which becomes dominant below $\sim 1$\,Hz and cannot be mitigated by improved isolation alone~\cite{Saulson:1984ap, Harms:2019dqi}. This sets an immutable low-frequency cutoff, excluding a variety of phenomena of deep astrophysical and cosmological interest~\cite{Christensen:2018iqi, Regimbau:2011rp}.  First, binaries of intermediate-mass black holes (IMBHs) with total masses exceeding $10^3\,M_\odot$ merge at frequencies below the terrestrial sensitivity band. Such systems, if detectable, would shed light on the poorly understood bridge between stellar-mass and supermassive BHs. Similarly, signatures of residual eccentricity---which encode formation histories of compact binaries in dense environments---manifest primarily at frequencies below a few hertz and will therefore remain hidden to CE and ET~\cite{Favata:2021vhw, Porter:2010mb}.

Another critical limitation concerns cosmology with dark sirens, i.e., GW sources lacking EM counterparts. Although CE and ET will identify thousands of compact binary mergers per year, their ability to localize such events to host galaxies will be limited~\cite{Chen:2018rjc, Borhanian:2022czq}. Only a handful of exquisitely localized dark sirens will be available annually, restricting their utility as precision cosmological probes.  Finally, searches for primordial or exotic stochastic GW backgrounds will be constrained by an irreducible astrophysical foreground from compact binaries~\cite{Zhou:2022nmt, Zhong:2024dss}. This astrophysical ``confusion noise” dominates ground-based detectors, masking the fainter cosmological signals that would offer insights into the physics of the early Universe, although notching and subtraction techniques could help mitigate foreground sources~\cite{Zhong:2025qno}. 

Detectors in space, away from the noisy earth, can probe the critical frequencies below $\sim 1$\,Hz. A deci-Hz observatory would bridge the frequency gap between LISA and CE/ET~\cite{Kawamura:2020pcg, Cutler:2009qv, Harms_2025}. Access to the 0.01–1\,Hz band would allow the detection of IMBH mergers, reveal eccentricity in compact binaries before circularization, provide much sharper sky localization for dark sirens, and probe primordial GW backgrounds beyond astrophysical foregrounds. Such an observatory would therefore open an entirely new discovery space, addressing precisely those science targets that remain beyond the reach of CE and ET.

\noindent
\textbf{Conclusions.}
CE and ET are poised to transform GW astronomy, offering unprecedented access to BHs and NSs across cosmic time, precision probes of dense matter, and novel opportunities in multi-messenger and fundamental physics. Yet important science targets remain beyond reach: the sub-Hz band is inaccessible due to Newtonian noise due to gravitational acceleration fluctuations, precluding the observation of very massive BHs and residual orbital eccentricity; only a handful of dark sirens can be localized well enough for cosmology each year; and searches for primordial stochastic backgrounds will be limited by astrophysical foregrounds. These frontiers point toward the need for future deci-Hz observatories, which could overcome these barriers and open an entirely new discovery space.

\subsection{Andrea Maselli: Einstein Telescope}

The Einstein Telescope (ET) is Europe’s flagship project for a 
next-generation GW detector. Recognized by the European Strategy Forum on Research Infrastructures as a priority facility, ET represents a major step forward in GW astronomy. The ET collaboration, formally established in 2022, has rapidly grown to nearly 2,000 members from 31 countries.

Two geometries are currently under consideration: a triangular 10-km–arm observatory consisting of three co-located, nested interferometers, each housing one interferometer optimized for low frequencies and one for high frequencies; and a network of two 15-km L-shaped nested interferometers located at separate sites in Europe.
The two-L configuration is expected to be moderately more costly, taking into account the tunnel diameters. In an L-shaped configuration, only two interferometers would be installed per tunnel, compared to four in the triangular configuration. Furthermore, the structural complexity of the triangular design poses significant challenges that could affect construction and increase overall building costs. Analyses in~\cite{Branchesi:2023mws} suggest that, for compact binaries, the two-L configuration may perform better than the triangular design, yielding up to a factor of $\sim$2–3 improvement in detection rates and parameter-reconstruction accuracy. Candidate host regions include Sardinia, the Euregio Meuse-Rhine, and Lusatia, with national governments committed to providing financial support if selected. A final decision on both geometry and site is expected by 2027.

Meanwhile, the ET collaboration is actively exploring the instrument’s scientific potential, with particular focus on how design choices will shape its capacity for discovery~\cite{ET:2019dnz,Branchesi:2023mws}. The recently published ET Blue Book by the Observational Science Board provides a comprehensive overview of the detector’s impact on fundamental physics, cosmology, stellar evolution, the astrophysics of compact objects, and the behavior of matter under extreme conditions~\cite{Abac:2025saz}.

The predicted sensitivity curve of ET shows an order-of-magnitude improvement over second-generation detectors. Crucially, ET will overcome the seismic noise wall that limits current interferometers below 10\,Hz, thanks to its underground location and the so-called xylophone configuration. Each ET detector will combine two interferometers: a cryogenic, low-power instrument optimized for low frequencies, and a room-temperature, high-power instrument for high frequencies. Access to frequencies below 10\,Hz will be essential for observing high-mass BH binaries and systems with large mass asymmetries. At the opposite end of the spectrum, enhanced sensitivity at kilohertz frequencies will enable the study of post-merger signals from compact binaries, allowing mode spectroscopy of both BHs and NSs.

Based on current models of compact-object evolution, ET is expected to detect on the order of $10^5$ binary BHs per year (about a hundred with signal-to-noise ratio $\sim$1000) and a comparable number of binary NSs (about a hundred with signal-to-noise ratio $\sim$100). Its detection horizon will extend to cosmological distances up to $z \sim 10^2$. Thanks to the low-frequency access, ET will also be sensitive to IMBHs with masses of order $10^4 M_\odot$. Together, these observations will enable hundreds of multi-messenger detections each year.

Low-frequency sensitivity will also allow ET to localize sources with high accuracy, which is crucial for fast and efficient follow-up of EM counterparts. ET is expected to detect $O(10^2)$ binary NS mergers per year with a sky localization of $\lesssim 100\, {\rm deg}^2$~\cite{Ronchini:2022gwk}. In a network configuration, with ET complemented by a CE–type detector, this number could increase to $O(10^3)$ binary NS mergers per year localized within $\lesssim 10\, {\rm deg}^2$.

Moreover, the ability to track long, slowly evolving low-frequency signals will enable ET to provide pre-merger alerts. For binary NSs, ET alone is expected to deliver $O(10)$ events per year localized within $<30\, {\rm deg}^2$ five minutes before merger. In combination with CE, this number could rise to $O(100)$ events with similar localization accuracy prior to plunge~\cite{Banerjee:2022gkv}. Together, ET and CE will generate a large number of well-localized triggers for EM follow-up. Given such volumes, it will be essential to develop data analysis strategies that prioritize events according to specific science goals.

Finally, synergies with detectors operating in different frequency bands, such as LISA and LGWA, will further enhance ET’s scientific reach. Multi-band observations will improve parameter estimation and expand astrophysical science cases, particularly those related to stellar evolution. For example, joint measurements across detectors will allow full coverage of the IMBH mass range and its distribution, opening a new observational window on their formation and growth~\cite{Abac:2025saz}.

\subsection{Ira Thorpe: LISA}

Ira Thorpe, NASA Project Scientist for the Laser Interferometer Space Antenna (LISA) mission~\cite{Colpi:2024hlh}, provided an update and overview. LISA's formal measurement band is $10^{-3}\,$dHz -- $10\,$ dHz, meaning that LISA will likely provide the first GW measurements in the deci-Hz band. LISA's sensitivity decreases towards the upper end of the band due to the large size of the constellation relative to the GW wavelength and a hard limit at $10\,$ dHz is set by the sampling rate.  

LISA's primary science objectives include the observation of the binary inspiral and merger of massive BHs ($10^3\,M_{\odot}\,\sim 10^7\,M_{\odot}$) out to extreme redshift ($z\gtrsim 30$); of the complex inspiral of stellar-mass compact objects around massive BHs in nuclear clusters; and of tens of thousands of compact binary systems in the Milky Way.  LISA is an example of the kind of ``big ideas" that are under discussion at this workshop.  Early work by Pete Bender and others dates back to the mid-1970s and it has taken several attempts to get the mission to where it is today, on track to launch in the mid-2030s.

LISA is really large, $2.5\,$Mkm on a side, which makes the constellation large enough for the Sun to fit inside it were the orbit to be heliocentric.  Scaling up the size of the detector proportionally scales the size of the displacements that are due to a given GW strain amplitude, so LISA's strain measurement problem is \emph{much} easier than LIGO's.  This is like the CE idea of scaling up LIGO, except it is by a factor of a million instead of ten.  Note that a deci-Hz mission based on the LISA concept moves in the other direction---as the arms shrink, the displacement measurement must get proportionally more precise in order to maintain the same strain sensitivity. Shot noise provides the fundamental limit and is straightforward to overcome, but there are other noises such as thermoelastic noise, geometric couplings from other degrees of freedom, and readout noise that need to be carefully considered. Many of these noise sources are intrinsically reduced when moving to higher frequencies, but the details will matter. 

A key feature of LISA is that it does not require absolute stability of the arms at the picometer level. This allows the spacecraft to orbit in free-fall with no active station keeping required to maintain the arm lengths.  The arm lengths will vary by up to one percent over the length of the mission, but the motion is on annual timescales, and so much below the measurement band of interest. The interferometric system is designed to accommodate slow drifts in arm length, range rates, and opening angles that result from these slow drifts. 

The drag-free control system effectively isolates the LISA test masses from external disturbances such as solar radiation pressure and micrometeoroids. The remaining disturbances on the test mass mainly result from the spacecraft itself and include residual gas pressure (mitigated by keeping the test mass in a low vacuum environment),  electrostatic couplings (mitigated by controlling the electric charge on the test mass with UV light), magnetic couplings (mitigated by selecting a test mass material with low magnetic susceptibility and high density), and Newtonian gravitational couplings (mitigated by carefully controlling the mass distribution, and thus temperatures, on the spacecraft).  The required free-fall performance for LISA was demonstrated by LISA Pathfinder~\cite{Armano:2018_lpf}.

A question prompted a discussion of the impact of having multiple space constellations operating simultaneously. In general, LISA does not require a second detector for anticoincidence because most of the anticipated binary-system sources have large signal-to-noise ratio and are in band for many cycles, which means the signal competes with many realizations of the detector noise, reducing the impact of rare noise excursions on detection. LISA also has two quasi-independent interferometers in its three arms, which can provide some level of anticoincidence, at least in the long-wavelength limit. While these approaches are sufficient for LISA's expected astrophysical sources, unambiguously detecting an unmodeled stochastic source, such as may arise from an inflationary cosmological GW signal, may prove challenging. A second detector in the LISA band would dramatically improve searches for unmodeled stochastic signals as well as improve localization for short-lived signals with clear benefits for multi-messenger studies. The combination of LISA and a deci-Hz detector would enable more multi-band GW astronomy.

In response to a question about calibration, it was pointed out that LISA does not require the same approach for calibration of GW strain response that terrestrial detectors like LIGO use; no active forces are applied along the GW measurement direction. Strain calibration comes from knowledge of the armlength and knowledge of the laser wavelength, both of which are known to better than 100\,ppm. This should be sufficient for sources which have a signal-to-noise ratio of up to $10^4$. 

Lastly a discussion of the impact of potential US budget cuts to NASA on the LISA mission emphasized that NASA is providing roughly $1/4$ -- $1/3$ of the funding and that ESA would likely be capable of absorbing the additional cost, albeit with a potential delay. No details are fixed.  It is noted that explicit funding for LISA was included in the Fiscal Year 2026 appropriations bill signed into law in January 2026, which significantly decreases the risk of near-term budget cuts to the program.  Teams in Europe and the US are continuing to work towards a 2035 launch date.

\subsection{Chiara Mingarelli: Pulsar Timing Arrays and targeted searches}

Following the recent PTA evidence for a nanohertz gravitational wave background (GWB;~\cite{NANOGrav:2023gor, EPTA:2023fyk, Reardon:2023gzh, Xu:2023wog, Miles:2024seg, NANOGrav:2024tnd}), the next milestones are the detection of individually resolvable supermassive black hole binaries (SMBHBs) and/or the anisotropy these induce in the GWB; see e.g.~\cite{Burke-Spolaor:2018bvk} and/or~\cite{2025NatAs...9..183M} for a recent, brief overview. Even under the most optimistic conditions, all-sky continuous GW searches are limited by poor sky localization, spanning  $29 - 241$~deg$^2$~\cite{Petrov:2024hec}. This motivates the development of targeted searches that incorporate EM priors on sky position, luminosity distance, and binary frequency~\cite{NANOGrav:2020lwu}. This approach improves strain sensitivity at specific sky locations and directly connects PTA science to multi-messenger follow-up.

The NANOGrav 15-year targeted searches program~\cite{Agarwal:2025cag} analyzed 114 active galactic nuclei (AGN) suspected of hosting SMBHBs. The vast majority of these AGN are based on the 111 periodic light curves identified in the Catalina Real-time Transient Survey~\cite{Graham:2015tba}.  Two additional candidates, PKS 2131-021 and PKS J0805-0111, from the Owens Valley Radio Observatory (OVRO). Furthermore, the study updated mass upper limits on galaxy 3C 66B, which was previously used as a pilot study~\cite{NANOGrav:2020lwu}, and found that their mass limits had started cutting into the EM-based mass estimates. 

Using fixed priors on sky location, luminosity distance, and frequency derived from EM data, the new NANOGrav targeted search reports typical strain upper limits improvements which are on average by a factor of $2.6\times$  relative to all-sky upper limits. Two targets exhibit weak Bayesian preference for a continuous-wave signal: SDSS J153636.22+044127.0 ( internally nicknamed ``Rohan'') and SDSS J072908.71+400836.6 (``Gondor''). These are not claimed as detections but are instead treated as two SMBHB candidates, and important case studies for developing a detection protocol.

A preliminary detection roadmap has been established based on a sequence of statistical and astrophysical consistency tests designed to evaluate the targeted candidates. These are first applied to SDSS J153636.22+044127.0 and SDSS J072908.71+400836.6, but are broadly applicable to AGN and quasar-based SMBHB candidates. These tests are summarized in Table 2 of~\cite{Agarwal:2025cag}, and include extended periodicity analyses using updated time-domain data, coherence tests to distinguish true GW signals from pulsar-dependent noise, dropout tests to verify individual pulsar support, population consistency checks, comparison with GWB-based anisotropy expectations, and separation from GWB systematics under both two different red noise models. The first is the common uncorrelated red noise (CURN) model, where the GWB manifests as a common red noise signal in the pulsars but has no Hellings and Downs cross-correlations, and the other red noise model has the Hellings and Downs cross-correlations expected for an isotropic GWB \cite{Hellings:1983fr}. 

Applied to J153636.22+044127.0 and SDSS J072908.71+400836.6, these tests yield mixed outcomes: some analyses marginally favor a coherent continuous-wave signal~\cite{Becsy:2025yzj}, while others indicate sensitivity to specific pulsars or model assumptions. No candidate satisfies all criteria required for a detection.

A central focus of ongoing work is the calculation of empirical trial factors associated with targeted searches, accounting for the number of sources explored. This effort is essential for establishing globally valid false alarm probabilities and will complement the  tests already performed. A simulation campaign is underway to calibrate significance levels and to complete a  detection checklist for future targeted searches. 

The emerging conclusion is that the key outcome of the targeted search effort is not a specific candidate but rather a structured detection framework that will enable future PTA discoveries. With EM-identified binary candidates providing potential host galaxy associations, the method also lays the groundwork for eventual LISA binary host galaxy identification, and eventually, standard siren measurements of $H_0$ if secure distances to the host galaxies are independently measured by GWs~\cite{Wang:2022oou}.

\subsection{Alessandra Buonanno: GW-Space 2050 Working Group}
\label{sec:GWSpace2050}

On June 11, 2021, the ESA’s Director of Science and the Science Program Committee (SPC) announced the next ESA Science Program, Voyage 2050, selecting three themes for the upcoming Large (L) missions. These themes were recommended in the Voyage 2050 Senior Committee report~\cite{ESAVoyage2050report:2021}, published in May 2021. The third L-theme is titled “New Physical Probes of the Early Universe”. The Voyage 2050 report emphasized: “How did the Universe begin? How did the first cosmic structures and BHs form and evolve? These are outstanding questions in fundamental physics and astrophysics, and we now have new astronomical messengers that can address them. Our recommendation is for a Large mission deploying GW detectors or precision microwave spectrometers to explore the early Universe at large redshifts. This theme follows the breakthrough science from Planck and the expected scientific return from LISA.” Additionally, the report stated: “Another approach that would shed a lot of light on the origin and growth of cosmic structures in the early Universe, say at $z  > 8$, is to open up a totally new spectral window for GW astronomy, either in the deci-Hz or micro-Hz (notably from the sub-micro-Hz to milli-Hz or from milli-Hz to Hz).” 

Following a recommendation from the LISA Consortium Executive Committee, the GW-Space 2050 Working Group was established in the summer of 2022 by a Steering Committee initially comprising Alessandra Buonanno, John Conklin, Michele Vallisneri, and Stefano Vitale. After initial discussions, quantitative work began in the fall of 2023. The goal is to produce a report or White Paper~\cite{GWSpace2050:report} that synthesizes the scientific case and technological feasibility of a potential GW-Space mission in the micro-Hz, milli-Hz, and/or deci-Hz frequency bands, as part of the ESA Voyage 2050 Program and future NASA initiatives. This document will support the upcoming assessment phase when ESA forms the Expert Committee to deliberate on the third L-theme. The GW-Space 2050 activities are sponsored by the Gravitational Wave International Committee (GWIC).

The GW-Space 2050 Working Group comprises a \textit{Science Group}, chaired by Alessandra Buonanno, Christopher Berry, Giulia Cusin, and Marta Volonteri, and an \textit{Instrument Design Group}, currently chaired by Jean-Baptiste Bayle, Olaf Hartwig, Daniele Vetrugno, and Stefano Vitale. Members of the GW-Space 2050 Working Group were recruited internationally, and the group currently includes approximately 60 actively contributing members. Since the White Papers~\cite{Baker:2019ync,Baibhav:2019rsa,Sesana:2019vho,Sedda:2019uro} submitted for the Voyage 2050 Program relied solely on straw-person noise curves for potential missions, it became a priority for the \textit{Instrument Group} to conduct a detailed study to identify technologically feasible space-based GW missions on the timescale of 2050. This effort aims to establish realistic mission concepts that can support robust and well-founded science cases. After initial discussions comparing laser and atom interferometer approaches, the group concluded—consistent with the recommendation of the Voyage 2050 report~\cite{ESAVoyage2050report:2021}---that a space mission based on atom interferometry is not technologically feasible on the 2050 timescale (see also Sec.~\ref{subsec:Hogan} by Jason Hogan for a review on the status of atom interferometry). Thus, the \textit{Instrument Group} has focused on laser interferometry. The current findings for the deci-Hz mission are summarized in Sec.~\ref{sec:Vetrugno} by Daniele Vetrugno. Results for the micro-Hz and milli-Hz have become available by early December 2025.

The \textit{Science Group} is organized into nine Subgroups, each led by two coordinators. Below, we provide a brief summary of key activities within these Subgroups, which are developing the scientific case for space-based GW missions in the micro-Hz, milli-Hz, and deci-Hz frequency bands.

The \textit{Data-Analysis Subgroup}, coordinated by Michael Katz and Sylvain Marsat, has extended the data-analysis tools originally developed for LISA to enable the study of GW sources in the micro-Hz and deci-Hz frequency bands. This effort includes the implementation of new detector orbits and time-delay interferometry response functions in both time and frequency domains, covering a range of source types—transient, monochromatic, and inspiraling. Bayesian inference and Fisher information matrix techniques, adapted from LISA analyses, have been further refined and applied by the \textit{Science Group} to produce robust and reliable scientific projections.

A deci-Hz detector will enable the discovery and detailed characterization of IMBH binaries with component masses in the range $10^2 - 10^3\,M_\odot$, as well as IMBHs in binaries with stellar-mass BHs, white dwarfs (WDs), or NSs---known as intermediate mass-ratio inspirals (IMRIs). While such sources are beginning to be detected by the LIGO-Virgo-KAGRA (LVK) Collaboration, their population will be enhanced by next-generation ground-based detectors such as CE and ET. A space-based deci-Hz detector will probe IMBHs at much higher redshifts and observe their signals over longer durations than ET or CE, allowing for precise inference of their properties---such as masses, spins, luminosity distances, and sky positions---across cosmic time. Furthermore, a deci-Hz mission will detect the progenitors of LVK-observed signals, providing critical insights into their formation pathways through measurements of orbital eccentricity effects.

The \textit{Stellar-Origin BH and IMBH Binary Subgroup} is generating population catalogs for BH binaries formed through both isolated and dynamical channels, extending to higher redshifts than previously available. These simulations employ the COSMIC~\cite{Breivik:2019lmt} and B-POP~\cite{Sedda:2021vjh} codes, incorporating more than a dozen metallicity values to better capture the diversity of stellar environments. For IMBHs, results from Population III star models are used to inform the evolution of early-generation stars at $z \geq 10$, enabling a more accurate representation of the formation and evolution of the first massive BHs in the Universe. 

A micro-Hz detector will observe a rich population of GW signals from massive BH binaries at high redshift. These observations will provide crucial insights into the formation of quasars observed at $z \geq 7$, clarifying how the first massive BHs assembled their mass---through EM accretion or GW-emitting mergers---across cosmic time. A micro-Hz detector will not only observe merging massive BH binaries like LISA, but several hundred massive BHs with $10^7-10^8 M_\odot$ at few hundreds years before merger up to $z \sim 8-10$. It will also unveil the dynamics of stars and compact objects  orbiting SgrA*, and characterize hundreds of stellar-origin BH binaries in our Galaxy. 
A micro-hertz detector will not only detect merging massive BH binaries, as planned for LISA, but also observe several hundred BH binaries in the range $10^7-10^8\,M_\odot$, several hundred years before merger, extending to redshifts of $z \sim 8-10$. It will further reveal the dynamics of stars and compact objects orbiting Sgr A*, and enable the characterization of hundreds of stellar-origin BH binaries within the Milky Way. 

The \textit{Massive Black Hole Binaries and Extreme Mass-Ratio Inspirals Subgroup,} coordinated by Elena Maria Rossi and Alberto Sesana, is assessing and quantifying the ability of the proposed missions to probe the cosmic epoch of massive BH formation at high redshift ($z > 8$). It is also identifying key scientific opportunities in the lower redshift Universe, including tidal disruption events. The predicted event rates and BH properties of single GW sources are strongly influenced by the unresolved astrophysical backgrounds---particularly the Galactic WD binary background, but also the massive BH binary background  in the micro-Hz band and the extragalactic WD binary background in the deci-Hz band. Current estimates, which may evolve as final noise curves are established, suggest that in the heavy-seed scenario, a micro-Hz detector could detect approximately 50\% more sources than LISA. This enhanced sensitivity arises from its ability to observe binaries during the prolonged inspiral phase, enabling access to higher-mass systems and significantly improving the accuracy of distance measurements and sky localization. In the light-seed scenario, a deci-Hz mission would detect roughly four times more binary systems than LISA, extending the observational reach to higher redshifts. Moreover, the improved signal-to-noise ratio and longer observation times in the deci-Hz band would lead to more precise parameter estimation, particularly for BH spins and mass ratios.

Deci-Hz and micro-Hz missions will enable unique synergies with EM observations. A deci-Hz band detector will localize bright standard sirens---particularly binary NSs and mixed neutron-star–black-hole binaries---with arcminute angular resolution, thanks to the prolonged observation times of one year or more. This extended detection window will allow EM facilities to be pre-positioned to observe the merger events days or even years in advance. Unique to the deci-Hz band is the capability to observe WD binary mergers, followed by an EM counterpart. These observations will provide critical insights into the progenitors of Type Ia supernovae---specifically, whether they arise from the merger of two white dwarfs or from accretion onto a WD from a main-sequence or giant companion. Detection of GWs from such events would strongly favor the double WD scenario. 

As discussed, the micro-Hz band will detect GWs from inspiraling massive BH binaries with masses in the range $ M \in [10^7 - 10^8]\,M_\odot $, when each BH is likely still embedded within an accretion disk in an AGN. This enables simultaneous multi-messenger studies of their evolution, accretion dynamics, and luminosity through coordinated UV/X-ray observations. Moreover, IMRIs and extreme-mass-ratio inspirals (EMRIs) will be observed at higher redshifts—predominantly in AGNs---rather than primarily in quiescent galactic nuclei as expected for LISA. Consequently, their GW signals may be accompanied by detectable X-ray counterparts, offering a powerful probe of BH growth in dense, active environments. The \textit{Multimessenger Astrophysics Subgroup}, coordinated by Tamara Bogdanovic and Zoltan Haiman, is identifying compelling science cases for different GW bandwidths that can be uniquely addressed through the synergy of GW and EM or fundamental particle observations.

Next, the \textit{Cosmography Subgroup}, coordinated by Alberto Mangiagli and Miguel Zumalacarregui, is quantifying constraints on cosmological parameters---including the Hubble-Lema\^itre constant $H_0$—using EMRIs and stellar-mass binaries in both bright and dark standard siren scenarios, which will be multi-band sources detectable by ground-based facilities operating in the 2050s. With a micro-Hz detector, there is potential to constrain $H_0$ at high redshift ($z > 8$) using massive BH binaries; however, the number of such sources may be insufficient, and identifying their host galaxies may remain challenging. A micro-Hz detector will also be capable of probing dark-matter-dominated halos with masses below $10^8\,M_\odot$ through gravitational lensing (notably diffraction) distortions imprinted on the GW signal.

Missions probing the high-redshift Universe will enable a much deeper characterization of the Galaxy’s invisible stellar content, thanks to the higher signal-to-noise ratios with which these sources will be observed. The \textit{Galactic Compact Binaries Subgroup}, coordinated by Valerya Korol and Sylvia Toonen, is actively pursuing this critical science. While LISA has a $4-7\%$ probability of detecting a GW signal from a Type Ia supernova---assuming it arises from the merger of WD binaries---a deci-Hz mission with a horizon extending to the Virgo Cluster could increase this probability to nearly $100\%$. These observations may  reduce systematic uncertainties in the use of Type Ia supernovae as standard candles for measuring the Hubble-Lema\^itre constant, with important implications for resolving the current tension between local and high-redshift measurements. Furthermore, compared to LISA, a deci-Hz mission is better suited for detecting mergers between WDs and NSs—although such events are likely rare. A micro-Hz mission, in contrast, offers a more substantial advantage over LISA, with approximately twice as many Galactic sources detected, significantly improved measurements of chirp mass and distance, and enhanced sky localization. These themes are also being investigated by the \textit{Astrophysical Stochastic Backgrounds Subgroup}, coordinated by Irina Dvorkin and Nikolaos Karnesis. Their goal is to refine predictions for both the Galactic and extragalactic WD binary backgrounds, while developing more efficient algorithms leveraging GPU acceleration. Given that unresolved astrophysical backgrounds (foregrounds) will obscure the detection of a primordial cosmological GW background from the early Universe, improving these predictions and computational methods is essential for isolating potential primordial signals at very high redshift. In particular, a deci-Hz mission is affected by the stellar-mass BH and binary NS backgrounds from LVK sources, which at frequencies $\lesssim 40$\,mHz are below the extragalactic WD astrophysical background. Astrophysical foregrounds could be discriminated from cosmological ones by differences in the shape and anisotropy of their spectrum.  

Beyond-LISA missions will offer significantly greater prospects than LISA to probe energy scales inaccessible to terrestrial particle accelerators and to CMB experiments, by enabling observations of the earliest moments of the Universe at redshifts far exceeding those probed by CMB spectral distortions. deci-Hz and micro-Hz missions will be capable of detecting GW stochastic backgrounds from cosmic inflation, depending on the tilt of the primordial power spectrum, thereby revealing previously inaccessible epochs---from the end of inflation to Big Bang nucleosynthesis---in which the Universe was dominated by neither vacuum energy nor radiation nor matter. These detectors may also uncover evidence of first-order phase transitions, such as the electroweak and QCD transitions, as the particle physics landscape evolved in the Universe with decreasing temperature and energy. Furthermore, stochastic backgrounds from cosmic and fundamental strings, which span multiple frequency decades, could be constrained 2 or 3 orders of magnitude more sensitively than with LISA. 

The \textit{Cosmology of the Early Universe Subgroup}, coordinated by Chiara Caprini and Ema Dimastrogiovanni, is investigating the scientific implications of detecting any of these stochastic GW backgrounds. The subgroup is studying the prospects for subtracting astrophysical foregrounds that overlap with cosmological signals, and assessing whether prior observations of these foregrounds by LISA and ground-based detectors can inform and improve foreground removal. They are developing quantitative figures of merit for early Universe GW sources and are working to quantify the ability to reconstruct the morphology of the underlying signals, enabling deeper insights into the physics of the primordial Universe.

With GW detectors targeting high redshift, GW signals from binary systems will be detected with significantly higher signal-to-noise ratios and at greatly enhanced rates than LISA and ground-based detectors. This will enable substantially more precise tests of general relativity. A deci-Hz mission combined with a ground-based detector will probe the strong equivalence principle using multi-band observations of stellar-mass BHs and NSs, as well as IMRIs---where smaller masses correspond to stronger spacetime curvature. BH spectroscopy of the remnant formed after merger will be performed with unprecedented precision, and the multipolar structure of the BH spacetime will be mapped in detail using EMRIs and IMRIs. These observations will allow for the detection of subtle imprints from a third compact body or from surrounding dark-matter environments. The \textit{General Relativity and Fundamental Physics Subgroup}, coordinated by Emanuele Berti and Andrea Maselli, is leveraging inspiral, merger, and ringdown signals from diverse binary populations to quantify the constraints on the nature of gravity and the properties of compact objects that deci-Hz and micro-Hz missions will be able to achieve. They are also investigating the possible dependence of tests of general relativity on the redshift. 

In summary, the GW-Space 2050 Working Group is synthesizing the scientific and technical foundations for a future space-based GW observatory, after LISA, in the micro-Hz, milli-Hz, and deci-Hz frequency bands (although the above discussion did not cover the milli-Hz band). This effort builds on the recommendations of the Voyage 2050 Senior Committee and aims to provide a comprehensive assessment of mission concepts, science capabilities, and technological readiness for inclusion in the upcoming ESA third-L mission selection process. The work integrates detailed modeling of astrophysical and cosmological source populations, GW data analysis, instrument design, and foreground characterization across multiple frequency bands. Key outcomes include projections of source detection rates, parameter-estimation accuracy, and constraints on cosmological and fundamental physics models. The findings will inform the development of mission concepts and support the evaluation of the third-L mission theme, “New Physical Probes of the Early Universe,” within the ESA Voyage 2050 program. The GW-Space 2050 White Paper~\cite{GWSpace2050:report}, expected to be released before summer 2026, will serve as a key reference for space agencies in shaping the next generation of GW missions.

\clearpage

\section{Monday afternoon -- deci-Hz GW science: astrophysical and cosmological sources (discussion panel)}

\noindent
{\em Panel:
Francesco Iacovelli, Marc Kamionkowski, Konstantinos Kritos, Cole Miller, Joe Silk, Emanuele Berti (Panel coordinator)}

\vspace{.5cm}

The science case for deci-Hz GW experiments is extremely rich (see e.g.~\cite{Mandel:2017pzd,Sedda:2019uro,2021hgwa.bookE..50I,Ajith:2024mie}). Our discussion session did not strive for completeness, but rather identified certain interesting classes of astrophysical and cosmological sources. The discussion slides focused predominantly on LGWA given the availability of a comprehensive White Paper~\cite{Ajith:2024mie}, but the sources are common to all detectors, although differences in sensitivity may ultimately yield different and complementary science.

\subsection{Astrophysical sources}

\subsubsection{Binary black holes} 

Observations by current ground-based detectors allow us to characterize the distribution of binary BHs in the Universe at relatively low masses. Estimates show that multi-band observations with GW observatories that are sensitive in other frequency bands can be extremely informative, improving the reconstruction of the source properties~\cite{Sesana:2016ljz,Toubiana:2022vpp,Buscicchio:2024asl,Wu:2025zhc} and allowing for new or improved tests of GR~\cite{Barausse:2016eii,Perkins:2020tra,Baker:2022eiz}. The prospects of multi-band detections with the LISA mission seem challenging because event rates are low, and because many of the observable signals take a relatively long time to sweep from the milli-Hz band of LISA to the frequencies of order a few Hz where they become observable by ground-based observatories~\cite{Cutler:2019krq,Gerosa:2019dbe,Toubiana:2022vpp,Buscicchio:2024asl}. A deci-Hz detector would instead be able to deliver more multi-band observations, with time delays of ${\cal O}(1\,{\rm yr})$ for BH binaries merging in the Hz band~\cite{Iacovelli:2025kwn}.
In particular, the long observational time over which dephasings can accumulate and become detectable imply that deci-Hz binary BH observations can yield stringent constraints on deviations from GR in the post-Newtonian (PN) coefficients, especially at low PN orders~\cite{LGWA:2020mma}. 

\subsubsection{Binary neutron stars}

Similarly to the binary BH case, a deci-Hz detector would be able to observe NS binaries well before their merger in the band of ground-based detectors. These signals can spend years inspiralling at frequencies in the deci-Hz band. This would provide early alerts for the sources and improved sky localization, increasing the odds of identifying an EM counterpart to the merger (see e.g.~\cite{Ajith:2024mie,Yelikar:2025jwh}).

\subsubsection{Intermediate-mass black holes} 

A key target for deci-Hz observations are IMBH binaries. As was the case for GW190521~\cite{LIGOScientific:2020iuh} and GW231123\_135430 (hereafter GW231123)~\cite{LIGOScientific:2025rsn}, binaries with masses higher than ${\cal O}(100\,{\rm M}_\odot)$ will spend only a few cycles in the band of current ground-based interferometers, and even future instruments like ET and CE will only be able to observe them for at most a few tens of seconds~\cite{Amaro-Seoane:2009vjl,Jani:2019ffg}. The deci-Hz band is optimal to observe IMBH signals, which can spend months to years inspiralling at those frequencies, resulting in a much more accurate determination of their parameters. A deci-Hz detector can break parameter estimation degeneracies when combined with ground-based observations, and it is optimal to characterize this new population of sources, thus shedding light on the role of hierarchical mergers in the formation and growth of supermassive BHs.

\subsubsection{Double white dwarfs} 
\label{sec:DWD}

Another interesting class of binary systems observable in the deci-Hz band are short-
period and merging double WDs~\cite{Benetti:2025rxe}. These binaries can be observed at signal-to-noise ratio $\sim 4$ out to $\sim 50$~Mpc with LGWA (see Figure~7 of~\cite{Ajith:2024mie}), assuming the maximum GW frequency $\sim 0.1$~Hz achievable with a binary of standard $M\approx 0.6~M_\odot$, $R\approx 10^4$~km WDs.  This is several times the horizon distance available to LISA for these sources~\cite{Ajith:2024mie}.  If supernovae type Ia arise from the merger of double WD binaries, then the estimated volumetric rate $\sim 2\times 10^{-5}~{\rm yr}^{-1}~{\rm Mpc}^{-3}$~\cite{2024A&A...689A.203P} would imply a WD merger rate of $\sim 10~{\rm yr}^{-1}$ out to 50~Mpc.  Moreover, the host galaxy will be determined with high confidence.  To see this we start by noting that Figure~9 of~\cite{Ajith:2024mie} shows that after roughly a year, a source at $\sim 0.1$~Hz will be localized to $\sim 10^{-3}~{\rm deg}^2$.  Using an estimated volumetric density of Milky Way Equivalent Galaxies (MWEG; this is a measure of the total stellar luminosity, divided by the stellar luminosity in the Milky Way) of $\sim 10^{-2}~{\rm Mpc}^{-3}$~\cite{2008ApJ...675.1459K}, there will be $\sim 5\times 10^3$ MWEG out to 50~Mpc.  There are $\sim 4\times 10^7$ patches of $10^{-3}~{\rm deg}^2$ in the whole sky, so the average patch will contain $\sim 10^{-4}$ MWEG, allowing for confident galaxy identifications.

Such detections will improve our understanding of double WD coalescence rates. This by itself will yield valuable information about the match, or discrepancy, with supernova~Ia rates.  They will also introduce a challenge and opportunity: at frequencies within a factor of several of the merger frequency, tidal effects will change the gravitational waveform. If new templates can be developed, they will inform us about the structure of the WDs, as well as their chirp mass and (possibly) the component masses.  Such  information will feed back into studies of supernovae~Ia and, as we discuss in Section~\ref{sec:WDmergers}, it can improve our understanding of these systems, and hence of cosmological estimates that rely on supernova~Ia physics.

\subsubsection{Intermediate mass-ratio inspirals} 

Intermediate-mass-ratio inspirals (IMRIs), i.e., binaries with mass ratios $q=m_1/m_2\gtrsim{\cal O}(10^3)$, will spend a large number of cycles in the deci-Hz GW band, allowing for precise parameter estimation and tests of GR~\cite{LGWA:2020mma,Ajith:2024mie}. 
Studies have predicted IMRI rates with a primary mass in the few hundreds of $M_\odot$ of at least a few per year in the local Universe~\cite{Mandel:2007hi}, with those from captures being the most eccentric.
It is expected that a significant number of IMRIs are formed dynamically in dense star clusters, where the central IMBH can ``swallow'' stellar-mass BHs from the surrounding environment~\cite{Kritos:2022non}.
These IMRIs will unambiguously probe properties of IMBHs, shedding light into their mass and spin distributions while these massive BH seeds are being assembled~\cite{Kritos:2024upo,Purohit:2024zkl}.

\subsubsection{Tidal disruption events}

When a star is captured by a massive BH and torn apart and stripped by its gravitational pull, a powerful burst of EM and GW radiation can be emitted~\cite{Sesana:2008zc}. In particular, WDs can be tidally disrupted by IMBHs~\cite{Maguire:2020lad}, and a new candidate has been identified in X-rays at redshift $z\sim1$~\cite{Li:2025mae}. The GW frequency of a WD in a circular orbit at the tidal radius around an IMBH is twice the Keplerian orbital frequency and scales as $\approx0.3\,{\rm Hz}(M_{\rm WD}/0.6\,M_\odot)^{1/2}(R_{\rm WD}/5000\,\rm km)^{-3/2}$. This GW frequency does not depend on the IMBH mass; however, for typical WD parameters, it falls in the deci-Hz range. This kind of signal would be detectable in the deci-Hz band for systems composed of an IMBH and a WD, and it would be an ideal multi-messenger target~\cite{Ajith:2024mie}. 
Specifically, since WDs are made up of degenerate material, they will explode similarly to a Type-Ia supernova when compressed deep within the tidal radius~\cite{Anninos:2018kzx}. The full tidal disruption of WDs produces a GW background detectable only by a deci-Hz observatory, with a signal-to-noise ratio of $\sim19$ over a 10-year period~\cite{Toscani:2025uar}.

\subsubsection{Eccentric binaries}

Binaries tend to circularize as they inspiral due to the emission of GWs: their eccentricity rapidly decreases with the orbital frequency, scaling like $e\sim f^{-19/18}\approx f^{-1}$~\cite{Peters:1963ux}. Thus, stellar-mass binaries that merge at frequencies $>1\,\rm Hz$ typically have much lower orbital eccentricities than they did during their early inspiral phase. 

High eccentricities are easier to measure, so low-frequency detectors such as LISA have better chances to measure nonzero eccentricity (see e.g.~\cite{Nishizawa:2016jji}). 
This is astrophysically relevant, because eccentricity measurements can be used to assess the relative contribution of different binary BH formation channels (see e.g.~\cite{Nishizawa:2016eza,Breivik:2016ddj,Zevin:2021rtf,Fumagalli:2024gko}). 
Binaries that are assembled dynamically in dense star cluster environments tend to have higher eccentricities relative to those that evolve from isolated massive binary stars (see e.g.~\cite{Kowalska:2010qg,Nishizawa:2016jji,Rodriguez:2018pss,Kritos:2022ggc}). In particular, more than $1\%$ of dynamical mergers have eccentricities $>0.1$ at $10\,\rm Hz$~\cite{Samsing:2017rat}, and AGN disks can also produce highly eccentric binaries~\cite{Samsing:2020tda}. Dynamical subchannels include BH-BH captures during single-single and binary-single interactions, Lidov-Kozai oscillations in hierarchical triples, and hardening of binaries through binary-single encounters.

Deci-Hz GW detectors will play a particularly important role in measuring their eccentricities -- and they will present us with new gravitational waveform modeling challenges~\cite{GilChoi:2022waq}. This is because many of the binaries that merge dynamically in star clusters--- especially those originating from gravitational captures and Lidov-Kozai mergers~\cite{Wen:2002km} -- actually {\em form} at deci-Hz frequencies, so they are not even observable at the mHz frequencies accessible with LISA~\cite{Samsing:2018isx}. Note in particular that captures between BHs in dense star clusters are expected to contribute significant numbers of eccentric mergers~\cite{OLeary:2008myb,Samsing:2019dtb}.
Numerical simulations of astrophysically motivated star cluster populations predict that about 15\% (33\%) of all dynamically assembled binaries will have eccentricities above 0.1 (0.01) at $0.1\,\rm Hz$~\cite{Kritos:2022ggc}, the vast majority of which are BH-BH captures and Lidov-Kozai mergers in BH triples.

\subsection{Cosmological sources}

Deci-Hz detectors present interesting opportunities for cosmology: 

\noindent
(i) Binary detections can be used for cosmography. Compared to the binaries that we are observing with current detectors, deci-Hz observatories would not only increase statistics, but they would also provide observations with good localization, which are helpful (e.g.) to implement  ``golden dark sirens'' techniques~\cite{Borhanian:2020vyr,Chen:2025qsl,Dang:2025vqx}; 

\noindent
(ii) They may identify individual events of cosmological origin, such as binaries of primordial BHs at high redshift and bursts of cosmic strings; 

\noindent
(iii) They could observe the SGWB generated by different processes in the early Universe, including inflation, first or higher order phase transitions (at energy scales higher than the ones accessible by LISA~\cite{Ajith:2024mie}), cosmic strings, and domain walls. Regarding SGWB detection, a key requirement is the availability of more than one data channel to perform a cross-correlation search~\cite{Maggiore:2007ulw,Romano:2016dpx}.

The detection of SGWBs of cosmological origin is contingent upon subtraction of both resolved astrophysical signals and of the SGWB generated by the unresolved ones. Depending on the frequency range of the detector, extragalactic unresolved DWDs can generate a relevant contribution to the background, but their distribution is still uncertain~\cite{Staelens:2023xjn,Boileau:2025jkv}. The subtraction of BH and NS binaries should instead represent a less severe problem~\cite{Cutler:2005qq}, thanks also to the reconstruction of their background achievable with ground-based detectors (see e.g.~\cite{Zhong:2025qno}), which could inform Bayesian source separation studies.

\subsection{Tests of general relativity}

Low-frequency, long-wavelength GWs observed by deci-Hz detectors provide a uniquely powerful laboratory for testing general relativity and probing possible extensions of gravity in the strong-field regime~\cite{Will:1971zzb,Berti:2015itd,Barack:2018yly,Yunes:2009ke}. By accessing compact binaries across a broad range of masses and evolutionary stages, deci-Hz observatories will explore a population of sources largely complementary to those observed by ground-based interferometers, enabling independent and synergistic tests of fundamental physics.

The inspiral signals observed in the deci-Hz band, characterized by the accumulation of thousands of GW cycles, make these detectors exceptionally sensitive to deviations from general relativity encoded in the PN expansion of the waveform. Agnostic frameworks that parametrize generic departures from general relativity across all PN orders are particularly well suited to this regime, as deci-Hz observations can place stringent constraints on pre-Newtonian (i.e., negative-PN) corrections that enter the dynamics before the leading quadrupole emission and are often associated with additional degrees of freedom or long-range modifications of gravity, providing potentially unambiguous signatures of new physics~\cite{Nair:2019iur}. 

At the same time, deci-Hz detectors significantly improve constraints on positive-PN corrections that dominate at higher frequencies, especially when combined with third-generation ground-based detectors such as ET or CE. This is because multi-band observations break degeneracies among waveform parameters and dramatically improve bounds on general relativity violations~\cite{Barausse:2016eii,Gnocchi:2019jzp,Perkins:2020tra}. 

The inspiral phase also enables precise measurements of the multipole moments of the compact objects, which constitute a powerful and complementary set of observables to test their nature and, in particular, to search for departures from the Kerr geometry expected for BHs in general relativity, as would arise in the presence of exotic compact objects~\cite{Krishnendu:2017shb, Cardoso:2019rvt,Bena:2020see,Vaglio:2023lrd}. 

Beyond the inspiral, deci-Hz observations will further allow high signal-to-noise ratio measurements of the quasinormal mode frequencies of intermediate-mass BH, enabling precision BH spectroscopy~\cite{Berti:2005ys,Meidam:2014jpa,Brito:2018rfr,Maselli:2019mjd,Capano:2020dix, Carullo:2021dui,Maselli:2023khq,Berti:2025hly}. The joint observation of the inspiral and ringdown across different frequency bands can yield independent measurements of the component masses and spins and of the remnant BH parameters, opening the door to stringent internal consistency tests of general relativity that directly link the binary dynamics to the properties of the spacetime of the merger remnant~\cite{Hughes:2004vw,Ghosh:2016qgn,Ghosh:2017gfp,Cabero:2019zyt}.

\clearpage

\section{Tuesday morning -- Overview of deci-Hz GW experiments}

\subsection{Lunar detectors}

\subsubsection{Jan Harms: LGWA}
The LGWA is proposed as an array of 4 stations, each equipped with 2 perpendicular horizontal inertial sensors~\cite{LGWA:2020mma,Ajith:2024mie}. The array serves to distinguish seismic background from GW signals through post-processing~\cite{Harms:2022awf}. The ideal array configuration must be small enough to analyze the seismic background coherently across the array, but large enough so that the seismic background has measurable differential phases or amplitudes between stations. The targeted lifetime of LGWA is 10 years, which can only be done with a nuclear-power device exploiting radioactive decays without fission. While nuclear power has been used in many space and planetary missions, it puts important constraints on mission aspects like transport, it is not easily available, and it is an important contribution to the mission cost.

It is known from Apollo seismic data that the Moon is an exceptionally quiet place with ground-vibration amplitudes 3 -- 4 orders of magnitude below terrestrial noise. The continuously present seismic background is not known today since the Apollo measurements were only able to set an upper limit. The contribution from meteoroid impacts is modeled to be at a few times 10\,fm/Hz$^{1/2}$ in the deci-Hz band. Sites at the lunar poles, especially in the permanently shadowed regions (PSRs) are expected to be seismically quieter. Equally important for GW detection is temperature stability. For a concept like LGWA, where deformations of the Moon induced by GWs are measured, temperature variations are expected to have a detrimental effect due to the thermal response of the ground, payload, or nearby structures like the lander. Flat PSRs are likely favored because the residual temperature variations on the relevant timescales will be caused by diffuse light from adjacent ground structures. Steep crater walls would cause relatively strong diffuse light to hit the ground of PSRs. Inside its polar PSRs, the Moon might well be one of the quietest planetary bodies in the solar system. Near-future missions like Chang'e 7~\cite{10.1093/nsr/nwad329} and the Farside Seismic Suite~\cite{10521223} will provide new insight into the seismic properties of the Moon and its internal structure. It is crucial to improve our understanding of the meteoritic hum and the Moon's internal structure for the planning of LGWA. 

The payload technologies and design are currently developed by groups in Europe and Asia. The core of the station is a pair of Watt's linkages whose displacements are read out by a compact laser interferometer~\cite{vanHeijningen:2023esw}. The position of the proof masses will be controlled by superconducting thin-film coil actuators. Operation of the station also requires a two-stage leveling system compatible with a cryogenic environment. The rough stage is to compensate initial ground slope up to 15\,deg. The fine stage needs to achieve about 20\,$\mu$ rad residual platform tilt. The superconducting coils require a ultra-low vibration sorption cooler unless suitable materials can be identified with critical temperature above the PSR ambient temperatures. An alternative readout technology is being studied using superconducting coils and SQUID amplifiers instead of the laser interferometer. The sensors will be tested in lunar emulators. One is being developed as part of the E-TEST surface facility in Liège. The other is being developed as part of the GEMINI facility 1.4\,km underground at LNGS. 

The proposal is for a single sensor array deployed in one PSR. Deployments at other locations, e.g., at the other lunar pole or at the Moon's equator, would enable additional studies~\cite{Harms_2025}. Searches of stochastic GW backgrounds become possible by correlating data acquired at different sites. A global network of stations makes it possible to measure the polarization of all GW signals including non-GR polarizations. The non-PSR deployments would still require a good temperature stability, which could be realized inside lava tubes or by burying stations into the regolith.

Connected to the temperature conditions at the LGWA site, an important component of the LGWA payload is the thermal management. The current payload design requires the sensor to remain below 15\,K to use superconducting coils as proof-mass actuators. This can be achieved with a two-stage sorption cooler. In this case, assuming most of the electronics and the laser to be at a central station, the main power consumption at the payload comes from the cooler itself, and two radiation panels with a total area of about 10\,m$^2$ will be needed if all the heat is to be radiated into space. Alternative superconductor materials are under investigation to operate the payload above 15\,K. This would make it possible to remove one of the sorption-cooler stations, and a single radiation panel would be needed with an area smaller than 2\,m$^2$. 

Given today's uncertainties of the lunar geophysical environment and its internal structure, a pathfinder mission is needed to validate the Moon as a platform for LGWA. This pathfinder was proposed to ESA under the name Soundcheck~\cite{Ajith:2024mie}, and it was selected by ESA into a reserve pool for lunar payloads waiting for a transport opportunity. Soundcheck will be deployed inside a PSR and its payload must therefore be compatible with operation at cryogenic temperatures. This provides an opportunity to develop new cutting-edge technologies for high-precision laser measurements, which might also find application in other areas, including active seismic isolation of the proposed ET. The targeted lifetime of Soundcheck is 2 months running on battery power. It will measure the temperature, magnetic field, and seismic background inside a PSR. 

When it comes to the observational capabilities of LGWA, an important aspect is that, similarly to LISA, sky localization is possible because of the rotation of the Moon and its motion around the Earth and Sun via phase and amplitude modulation of long-lasting GW signals. Sub-square-degree localization of all solar-mass binaries will be possible~\cite{Cozzumbo:2023gzs}. The observation band is constrained by instrument noise (especially suspension thermal noise) towards low frequencies and by the decreasing lunar GW response towards high frequencies, with peak sensitivity in the deci-Hz band.

The science output of LGWA depends on our ability to accurately calibrate LGWA data. The main challenge is to understand the response of the Moon to GWs. The model needs to be site dependent, and it can be improved with time by observing moonquakes. An alternative calibration method is to use pre-calibrated signals observed by other GW detectors, like future terrestrial detectors. Solar-mass binaries consisting of BHs or NSs, as well as IMBH binaries, could be used for such purpose. Observing enough of these binaries should map out the response of the Moon in dependence on the polarization of the GW and its propagation direction. This calibration can then be used for other types of signals only observed by LGWA, like the merger of WD binaries. One could turn this method around to infer the internal structure of the Moon from the response measurement in a tomography-type analysis.

An open problem is the distribution of meteorite impacts. Observations of impact flashes from Earth indicate that the polar regions are less affected, which is also supported by models, but more detailed studies are needed to confirm these findings. Magnetic fluctuations at the lunar surface are poorly understood, but this is an important parameter to know for LGWA because of the sensors susceptibility to magnetic fields. Radiation could charge the test mass, which might cause additional noise in the measurement, although the PSR should provide a certain level of protection against all radiation.

Synergies with other missions and experiments include any geophysical experiments on the Moon and any GW detector observing in frequency bands adjacent to the deci-Hz band, like LISA-type detectors or terrestrial detectors. 

The LGWA collaboration has more than 200 members today with working groups on GW science and multimessenger astronomy, lunar science and site studies, and instrument science. LGWA is a ``lid that fits on many pots'', i.e., white papers were submitted to ESA and NASA calls addressing polar as well as non-polar opportunities, robotic and human deployments~\cite{Harms_ESA,Harms_Artemis,Harms_BPS_research,Harms_2025}. The Soundcheck deployment could be achieved by 2030. Once the Soundcheck data are analyzed, the LGWA configuration can be finalized, and an LGWA deployment by 2040 is conceivable. The first step is to bring the integrated Soundcheck system to TRL6 in the coming few years.

\subsubsection{Karan Jani, V\"olker Quetschke, James Trippe: LILA}

The Laser Interferometer Lunar Antenna (LILA) is a proposed U.S.-led initiative to deploy a GW observatory on the lunar surface, designed to observe the mid-band frequency spectrum (0.1--10\,Hz)~\cite{Jani:2025uaz}. This frequency window bridges the gap between terrestrial detectors (limited by seismic and Newtonian noise below 10\,Hz) and future space-based detectors like LISA (limited by arm-length to milli-Hz frequencies)~\cite{Jani:2020gnz}. The LILA concept leverages the Moon's unique environmental advantages: a seismic background estimated to be orders of magnitude lower than Earth's, a natural ultra-high vacuum that surpasses terrestrial vacuum tubes, and a lack of atmospheric or acoustic noise~\cite{Cozzumbo:2023gzs}.

The mission is structured into a phased deployment strategy aligned with NASA's Commercial Lunar Payload Services (CLPS) and the Artemis programs:

\begin{itemize}
    \item \textbf{LILA-Pioneer (near-term):} Scheduled for the early 2030s, this phase consists of a long-baseline (3--5 km) optical interferometer deployed via CLPS landers and Lunar Terrain Vehicles~\cite{Creighton:2025kth}. Unlike terrestrial detectors, LILA-Pioneer operates as a non-suspended ``strainmeter'' where optical components are anchored directly to the regolith. This design exploits the Moon's mechanical response to GWs in two regimes: the inertial regime, where the surface acts like free masses, and the mechanical regime, where the detector utilizes the resonant amplification of lunar normal modes to achieve strain sensitivities competitive with orbital detectors~\cite{Majstorovic:2024jvl}.
    
    \item \textbf{LILA-Horizon (long-term):} Envisioned for the mid-2030s with human deployment support (Artemis IV+), this phase expands to a triangular 40--50 km observatory~\cite{Shapiro:2025oqa}. To reach broadband sensitivity targets across 0.1-10\,Hz, LILA-Horizon will incorporate active seismic isolation and suspended test masses -- specifically utilizing 100 kg masses supported by silica fiber suspensions to manage thermal noise.
\end{itemize}

LILA's scientific potential extends beyond standard GW astronomy. In the mid-band, it will detect binary NS inspirals and double WD mergers days to months before merger, providing early-warning alerts with $\sim 1$ arcmin$^2$ sky localization~\cite{2025ApJ...992...16P, Yelikar:2025jwh}. It will also conduct a cosmological survey of IMBHs ($10^2$--$10^6 M_{\odot}$) to high redshifts~\cite{Greene:2019vlv}. Crucially, LILA acts as a geophysical instrument; the strainmeter configuration is capable of detecting lunar normal modes with high signal-to-noise ratios, potentially including the Slichter mode of the inner core, thereby constraining the Moon's deep internal structure~\cite{Panning:2025jcx}.

Current technology development is coordinated through the International LILA Consortium and focuses on raising the Technology Readiness Level (TRL) of the laser subsystem. Key efforts include the development of a compact, fiber-based 1064 nm Master Oscillator Fiber Amplifier (MOFA) capable of withstanding the lunar thermal and radiation environment. Prototypes are undergoing validation in cryogenic vacuum chambers (Space Simulators) to simulate lunar conditions. The roadmap includes the development of the LILA-Pioneer Instrument Prototype (LILA-PIP) to validate the noise budget and system architecture required to secure a flight manifest~\cite{Jani:2025uaz}.

\subsection{Space-based detectors} 

\subsubsection{Kentaro Komori: DECIGO}

The Deci-hertz Interferometer Gravitational-wave Observatory (DECIGO)~\cite{Kawamura:2020pcg} is a planned space-based mission designed to open the deci-Hz frequency band of GW astronomy. This unexplored frequency range bridges the gap between ground-based detectors, sensitive above a few Hz, and space-based detectors such as LISA, targeting the mHz band. By operating in this middle ground, DECIGO aims to probe astrophysical and cosmological phenomena inaccessible to other observatories. Key science goals include detecting the stochastic GW background from inflation, observing the formation and evolution of compact binaries, and providing early warning signals for sources later observable by ground-based detectors. To achieve these aims, DECIGO will deploy multiple triangular constellations of spacecraft in heliocentric orbit, with arm lengths of 1,000 km and laser interferometer with optical cavities with $10^{-18}\,\mathrm{m/\sqrt{Hz}}$ precision. The detector will use large free-floating test masses and ultra-stable laser interferometer to reach strain sensitivities of order $10^{-24}\,\mathrm{/\sqrt{Hz}}$, enabling unprecedented access to the deci-Hz regime.  

Realizing the DECIGO mission requires overcoming several technical challenges. First, the inter-satellite laser interferometer must be controlled in all six degrees of freedom with extreme precision, a task complicated by the long arm length and the need for stable formation flying. Second, the large test masses pose major difficulties. B-DECIGO, the pathfinder mission, will employ 30\,kg test masses, while the full DECIGO will scale up to 200\,kg. Achieving drag-free control for such massive proof masses is unprecedented, demanding actuators with high authority yet vanishingly low noise. A third critical hurdle lies in meeting the force noise requirement of $10^{-16}\,\mathrm{N/\sqrt{Hz}}$. This translates into maintaining vacuum levels below $10^{-8}\,\mathrm{Pa}$, two orders of magnitude beyond what was achieved by LISA Pathfinder, as well as suppressing actuator noise to levels nine orders of magnitude smaller than the required DC actuation range of $10^{-7}\,\mathrm{N}$. Additional sources of disturbance, including charging noise, magnetic interactions, and plasma effects, remain insufficiently characterized but could pose significant risks. Each of these factors highlights the technological leap necessary to bring DECIGO from concept to reality.  

To bridge this gap, the Space Interferometer Laboratory Voyaging towards Innovative Applications (SILVIA)~\cite{Ito:2025izr} has been proposed as a near-term demonstrator mission. SILVIA consists of a three-spacecraft formation in low Earth orbit, separated by 100\,m, designed to validate ultra-precise formation flying at a manageable scale. The mission integrates spacecraft sensors, laser interferometry, and low-thrust, low-noise micro-propulsion to achieve sub-micrometer precision in both relative distance and orientation control. Its optical configuration employs an asymmetric Michelson interferometer, in which one arm extends directly after the beam splitter while the other arm spans the 100\,m separation, enabling a sensitive and compact design. Operating in a near-circular low Earth orbit minimizes relative perturbations while keeping collision risk low, providing a cost-effective and practical testbed. By addressing the integration of measurement and control technologies in real-time, SILVIA directly targets the key technological bottlenecks facing DECIGO. Beyond this role, SILVIA also opens opportunities for Earth science, for example through precision measurements of Earth’s gravity field and atmospheric perturbations using the same interferometric and formation-flying technologies. Moreover, its outcomes will extend beyond GW detection, supporting future missions such as the Large Interferometer for Exoplanets (LIFE)~\cite{2022A&A...664A..21Q}, aimed at imaging terrestrial exoplanets. SILVIA thus provides an essential stepping stone, enabling the maturation of core technologies that will realise the next generation of high-precision space observatories.

\subsubsection{Daniele Vetrugno: GW-Space 2050}
\label{sec:Vetrugno}

In response to ESA's Voyage 2050 call for new Large missions, the GW community submitted several White
Papers addressing different target frequency bands (see Section~\ref{sec:GWSpace2050}). 
One of the White Papers~\cite{Sedda:2019uro} advocated for a GW mission in the deci-Hz band. 
The paper analyzed several proposed missions or mission scenarios, including ALIA, DECIGO, atomic clocks, and 
two LISA-like deci-Hz observatories (DOs): DO-conservative and DO-optimal.
The GW-Space 2050 group, formed after the ESA call to analyze the 
feasibility and the science case of these missions, has focused on the LISA-like ones. For clarity, the LISA-like missions presented in the White Paper have been renamed DO-WP (DO-WP 
Conservative, DO-WP Optimal), while the new mission profiles produced by the present study are called 
DO-IT (for ``Improved Technology'', and as an encouragement to ``just do it!'').

The noise assumptions presented in the White Paper were not completely justified. They were based on 
illustrative sensitivity curves derived from simplified performance budgets, assuming shot noise and  
acceleration noise as the only limiting factors, without including the so-called technical noises.
The GW-Space 2050 Instrumental design team has reassessed the feasibility of those sensitivity curves by 
considering three representative scenarios: DO-IT Conservative, DO-IT Baseline, and DO-IT Challenging.

The baseline assumption for the DO-IT missions is a triangular LISA-like cartwheel constellation of 0.1\,Gm 
arm-length. 
This implies more stable orbits than LISA (0.2\,m/s of maximum relative 
velocity between the spacecraft, instead of 8\,m/s for LISA) with a significant impact on heterodyne frequency and TDI 
noise.
The DO-IT Conservative mission assumes 10\,W of laser power, while the 
Baseline and Challenging missions assume 20\,W. As a 
reference, the power in the DO-WP was 10\,W in the Conservative case, and 30\,W in the Optimal case.
Three options are considered for the Telescope diameter: 0.3\,m 
(Conservative), 0.5\,m (Baseline), and 1\,m (Challenging).
A first criticality of the DO-WP concept was indeed the assumed telescope 
diameter. 
The DO-WP assumed a 1\,m or 2\,m telescope. However, considering cost 
related to mass and volume of such 
big telescopes, already 1\,m appears highly challenging. 
The Conservative configuration assumes the same laser wavelength as LISA, 
1064\,nm, while the Baseline and Challenging options considers 532\,nm.

Starting from the LISA performance budget, the study evaluated all the 
known noise sources -- i.e. Long 
arm IFO noise, Test mass interferometer (TM IFO) and reference 
interferometer (REF IFO) noise, TDI residual noise, 
acceleration noise -- adapting them to the new configurations to assess 
what may be achievable in 15-20 
years.
It must be noted that the DO-WP sensitivity curve assumed an improvement 
in displacement noise in the 
deci-Hz band relative to LISA of $6\times10^{-4}$ (DO-WP Conservative) and 
$8\times10^{-5}$ (DO-WP Optimal).
Including technical noises, such improvements appear out of reach.

\noindent
\textit{Long arm IFO noise.} Shot noise as well as chain noise (the 
electronic noise from the photodetector to 
the phasemeter) can be reduced by increasing the transmitted power (more 
powerful laser, 
shorter armlength, larger telescopes). However, not all of those noises can 
be equally reduced by the new observatory configurations.
A key aspect is phasemeter noise: more laser power and more stable orbits 
could slightly help by reducing the need of voltage amplification. 
Experts in the field estimate that 0.1\,pm/$\sqrt{\mathrm{Hz}}$ is realistically attainable with further 
development. A more ambitious goal of 0.05\,pm/$\sqrt{\mathrm{Hz}}$ could 
be set, though this would require new technology that is not yet 
available.
Other critical aspects include stray light noise (in particular telescope back-scatter), which may be mitigated using 
frequency swapping configurations and temperature stability of 1\,$\mu$K/$\sqrt{\mathrm{Hz}}$ (LISA Pathfinder achieved 
10 $\mu$K/$\sqrt{\mathrm{Hz}}$, but performance should be easier at 0.1\,Hz); and TTL residuals, which are assumed to be correctable to below the 
shot-noise level -- a strong assumption, but the same one adopted for LISA.

\noindent
\textit{Test mass IFO noise.} LISA Pathfinder demonstrated a 35 fm/$\sqrt{\mathrm{Hz}}$ local IFO noise level.
For future missions, full axes IFOs will be needed for calibration purpose. This would allow local measurement 
of low-frequency TM noise, otherwise not directly measurable as in LISA, providing a proxy for acceleration 
noise at low frequency. It has been pointed out in the discussion that this is important for the SGWB. 
However, the need for instruments that are non-colocated to claim measurement of GWB remains.

\noindent
\textit{Reference IFO noise.} This noise source is assumed to be negligible, provided a major change in 
the spacecraft design: a single rigid optical bench (OB) with in-field pointing, eliminating the need of 
MOSAs. This would require significant technology development. In the discussion, it has been 
highlighted that single OB was actually studied by Airbus in the past.

\noindent
\textit{Acceleration noise.} One order of magnitude improvement in Brownian noise is required, but it
appears feasible, together with a higher bandwidth spacecraft controller.

\noindent
\textit{Laser wavelength.} Switching to laser light of 532\,nm (rather than 1064 nm, as in LISA) improves noise and sensitivity and helps mitigate plasma 
scintillation noise, but it also adds complexity.

Taking all these considerations into account, the DO-WP configurations show a degradation by about two orders 
of magnitude relative to their assumed displacement improvement over LISA.
For the DO-IT configuration, the study finds improvements with respect to LISA of $1.8\times10^{-2}$ for DO-IT Conservative, 
$3.5\times10^{-3}$ for DO-IT Baseline, and $6.8\times10^{-4}$ for DO-IT Challenging.  
The Challenging scenario overlaps almost completely with DO-WP Conservative, but it is already considered 
unrealistic due to the prohibitive cost of a 1\,m telescope.

The advantage of a LISA-like approach is clear: the mission profiles build upon a well-studied concept, 
allowing enormous benefit from LISA’s heritage. On the other hand, since LISA has not yet flown, its actual 
feasibility remains uncertain.

During the discussions, it became clear that ESA’s timeline for new Announcements of Opportunity will not wait for the LISA results before deciding on the 
next mission. There was also some consideration of modifying the arm length. However, this comes with trade-offs: shorter arms reduce sensitivity, while 
longer arms become limited by the WD foreground noise.

Near-future work will focus on justifying the DO-IT performance budget for the final GW-Space 2050 report or White Paper~\cite{GWSpace2050:report}, and on applying a similar approach 
to the microHz mission design. \textit{Just DO-IT!}

\section{Tuesday afternoon}

\subsection{Atom interferometry}

\subsubsection{Jason Hogan: MAGIS}
\label{subsec:Hogan}

Atomic sensors are an emerging GW detection approach that offers a promising route to the deci-Hz band. Freely falling neutral atoms can be well isolated from non-gravitational forces, allowing for an excellent realization of a gravitational ``test particle''.  Atoms can also serve as a pristine phase reference, with the best optical lattice clocks in the world now demonstrating fractional frequency stability below $10^{-18}$~\cite{2024PhRvL.133b3401A}.  Atomic clocks~\cite{Kolkowitz:2016wyg} and atom interferometers~\cite{Hogan:2015xla} are closely related techniques~\cite{Norcia:2017vwu}, with each offering different advantages for GW detection.

The atom interferometer GW detector concept uses freely falling atoms as inertial references on each end of a long measurement baseline, with discrete laser pulses propagating back and forth along the baseline that both manipulate the atomic state and transfer momentum to the atoms.  Using a sequence of laser pulses, the atomic wavefunction can be split, redirected, and ultimately recombined to realize a Mach-Zehnder interferometer geometry.  By making use of the same ultra-narrow transitions used in atomic clocks~\cite{Graham:2012sy}, the phase shift of each ``clock'' atom interferometer depends on the time the atom spends in the excited state, which is proportional to the light travel time across the baseline. By comparing the phase of these atom interferometers on each end of the baseline, the resulting gradiometer signal is sensitive to variations of the light travel time caused by GWs, and many common sources of technical noise arising from the laser cancel as a common mode.  In this clock gradiometer design, the atoms act as a kind of ``active'' proof mass, simultaneously serving both as inertial reference points and as a phase reference, avoiding the need for multiple measurement baselines to cancel laser noise.  Clock atom interferometers in this configuration offer sensitivity to both GWs and ultralight dark matter~\cite{Graham:2012sy,Arvanitaki:2016fyj}, and these signals can be distinguished thanks to their unique angular dependence (scalar versus tensor) and frequency evolution.

Several approaches can be used to enhance the sensitivity to GWs. Large momentum transfer (LMT) atom optics coherently split the atomic wavepackets with many sequential light pulses, increasing the separation between the interferometer arms.  Since each pulse in an LMT atom optics sequence samples the light travel time across the baseline, the use of LMT enhancement is analogous to the use of Fabry-Perot resonators in the arms of LIGO~\cite{MAGIS-100:2021etm}.  Proof-of-concept experiments have demonstrated over $400~\hbar k$ momentum transfer, which implies a factor of 400 enhancement in sensitivity~\cite{Wilkason:2022yej}.  Resonant sequences, consisting of multi-loop atom interferometers tuned to a particular frequency band, provide enhanced (narrow-band) response at a target frequency and are closely related to dynamical decoupling sequences used in quantum information science~\cite{Graham:2016plp}.  State-of-the-art demonstrations have recently realized phase amplification over hundreds of loops~\cite{Wang:2024puy}.

Several large-scale experimental efforts are advancing this field. There are 10-meter-scale atom drop towers at Stanford~\cite{Dickerson:2013ykr}, Hannover~\cite{Schilling:2020pka}, and Wuhan~\cite{Zhou:2011gmi}.  The MAGIS-100 experiment~\cite{MAGIS-100:2021etm} at Fermilab will realize a 100\,m vertical baseline strontium atom interferometer and gradiometer, and is scheduled for commissioning in 2027.  Initial sensitivity is expected to be at the $10^{-14}/\sqrt{\text{Hz}}$ level, with improvements of 10 to 100 anticipated during the first few years of operation.  MAGIS-100 will serve as a prototype deci-Hz GW detector, while also probing ultralight dark matter and demonstrating extreme quantum superpositions with meter-scale separations and multi-second durations. Other significant efforts are also already underway around the world, including MIGA~\cite{2018NatSR...814064C} in France, ZAIGA~\cite{2020IJMPD..2940005Z} in China, and AION~\cite{Badurina:2019hst} in the UK.  Looking to the future, an annual international workshop~\cite{Abdalla:2024sst} on Terrestrial Very-Long-Baseline Atom Interferometry (TVLBAI) has begun, and an international proto-collaboration has formed to study the prospects for kilometer-scale terrestrial atom interferometric detectors.

Space-based detectors based on the MAGIS concept have also been studied, where a pair of satellites are separated by a single measurement baseline. Each satellite contains a cold atom ensemble, with common laser light propagating between the two spacecraft.  For long baselines, heterodyne laser links can be used to phase lock incoming reference light to strong local oscillator lasers in each satellite that then interrogate the atomic sample, maintaining laser noise rejection~\cite{Hogan:2015xla}.  In this scheme, the desired measurement between the atoms in each satellite is effectively synthesized by combining a spacecraft-to-spacecraft measurement (via the reference light) plus two local spacecraft-to-atom measurements (via local oscillator lasers), analogous to the approach used in LISA.  Such detectors can operate in either broadband or resonant mode, providing sensitivity across the full deci-Hz band and avoiding limits imposed by Newtonian gravitational acceleration fluctuations for terrestrial detectors in this band.  There have also been proposals to use optical lattice clocks to detect GWs~\cite{Kolkowitz:2016wyg}.  In the clock-based approach, the atoms still serve as the phase reference, allowing common mode rejection of laser noise over a single measurement baseline.  However, in an optical lattice clock the atoms are rigidly trapped in a standing wave of light connected to the rest of the spacecraft.  Therefore the atoms no longer can serve as an inertial reference, requiring a LISA-like drag-free control system to avoid vibration noise. This tradeoff comes with the advantage that the wavefunction of the trapped atoms no longer spreads out like in the freely-falling atom interferometer approach, avoiding some technical challenges and some potential sources of noise arising from variations across the wavefunction.

Achieving scientifically relevant sensitivity requires substantial progress in three areas. For terrestrial efforts, additional improvements in LMT atom optics are essential, with simulations suggesting that enhancements of more than four orders of magnitude are feasible using optimized 100\,W-class laser systems and composite pulse techniques. Atom sources with higher flux are also needed.  There has been progress developing continuous sources of ultra-cold strontium~\cite{2020arXiv201207605C}, as well as rapid evaporative cooling techniques using dynamically shaped optical traps~\cite{2016PhRvA..93d3403R}, which could be part of a solution to increase flux by the factor of $\sim 100$ needed to reach the most ambitious targets~\cite{MAGIS-100:2021etm}.  Finally, quantum entanglement of the atoms can be used to further reduce phase noise below the standard quantum limit.  Spin squeezing has already been demonstrated at the 20\,dB level in rubidium~\cite{Hosten:2016yho}, and new squeezing systems based on strontium are under development.  High flux source development and spin-squeezing are particularly necessary for terrestrial detectors because of the much shorter measurement baseline compared to space-based configurations, but all detectors can benefit from these improvements.

Atom interferometry provides a powerful and fundamentally new approach to GW astronomy in the deci-Hz band. Active international efforts, coupled with near-term prototypes such as MAGIS-100, are rapidly advancing the required technology. Future large-scale terrestrial and space-based missions can open a new observational window, with the potential for transformative discoveries in astrophysics~\cite{MAGIS-100:2021etm} and cosmology~\cite{Graham:2016plp}.

\subsection{deci-Hz experiments and science -- discussion panel}

\noindent
{\em Panel: Jason Hogan, Kentaro Komori, Daniele Vetrugno, Jan Harms, Volker Quetschke, Joe Silk, Francesco Iacovelli, David Shoemaker (Panel coordinator)}

\vspace{.5cm}

The panel discussion was oriented around a few questions which intended to probe the challenges involved in pursuing the various proposed missions for exploring the deci-Hz domain. 

{\bf What’s the most fragile element of the science case?}

For all the deci-Hz missions, the search for a primordial GW background is both the most compelling and the most challenging to realize. The risk for science lies in the combination of the uncertainty in the level of a primordial GW background (which is the key desired science goal for DECIGO and an important one for both LILA and LGWA), and the uncertain level of the deci-Hz GW astrophysical background may mask the primordial background. Further measurements with terrestrial detectors and then LISA will enable a better estimate of the astrophysical background, reducing one risk. Further theoretical work and potentially discoveries of, e.g., sub-solar mass BHs can refine estimates for the primordial background.  

SGWBs are always challenging. It is hard to produce convincing evidence of the detection of a background, because there are always some instrumental noise correlations. Distinguishing between an astrophysical and a cosmological background is also hard. In order to lower the astrophysical foreground, it would be necessary to resolve a lot of individual CBC signals. LGWA and LILA might not be sensitive enough to do that.

For both LGWA and LILA, one can also make use of ground-based detectors as a complement. For instance, one can resolve more individual CBC signals via multiband, or can constrain the CBC population with the ground-based detectors, then use such information to put tight priors on the astrophysical background in LGWA and disentangle it from a cosmological background.

{\bf What sets the timeline for the proposed missions?}

Moving to space is key for all the deci-Hz missions to move away from the gravitational acceleration limit on the earth's surface, with both free-flyers and Moon-based systems as potential implementations. 

Atom interferometry is very promising for a lot of very high precision metrology tasks, and GW detectors are one such application. There are many reasons to launch atom interferometers in space within a few years, regardless of the GW application. Atom interferometers for GW observation are at a lower technical readiness level than interferometers or seismometers (although promising ultimately potentially better sensitivity). The group noted that it would be very helpful for the  atom interferometry community to put together some concrete figure of merit for the characteristic strain, not just projections; the lack of such a baseline has made it difficult to consider in roadmap efforts to date. The community is on a good trajectory to get such a strain curve for a 100\,m atom interferometer in 2-3 years, and perhaps even earlier for a 10\,m interferometer. 

A challenge for GW science is that it requires a huge technological and monetary investment in order to get to a first measurement in a frequency domain requiring new infrastructure and technologies. For instance, LISA Pathfinder was just a “technology demonstrator”; it didn’t make any GW detection. There’s however an intrinsic difference between the initial prototypes for terrestrial detectors (e.g, Garching 30\,m, Caltech 40\,m)  and LISA Pathfinder, where people started to develop and run the first coalescing binary searches, even though there was no expectation they would detect anything. LISA Pathfinder was instead built in a way that it could not detect any GW signal. That’s different from established fields in astronomy, where even preliminary instruments can make some meaningful science measurements. Differences in funding for Europe and the US are significant as well. ESA seems to be somewhat unaffected by political changes. That doesn’t seem to be the case in the US, and this is for instance what caused the cancellation of JWST pathfinder. It will be very interesting to see how China chooses to proceed, and any consequences for the European and US approaches. 

For LISA (as for many other space missions), an important element is that it is planned to carry out the mission with no upgrades or repairs. This leads to long and expensive development schedules to give a high confidence of success, and in the case of LISA, a pathfinder mission. Lunar detectors offer an alternative approach where maintenance, repair, and incremental improvements can be woven into the planning from an early stage. Indeed, some of the the interest for some nations to support the lunar detectors is to give astronauts valuable roles and experience. A good analogy for this is science at the south pole, which was developed in part because the US had political interests there.

If the ambitious plans are realized of nations to populate the Moon and enable complex missions, the timeline for Moon missions to get started may be sooner than for any free-flyer missions. One could ask if investments in Moon missions with their potentially limited sensitivity or bandwidth could delay or eliminate funding for a free-flyer deci-Hz mission. The GW community may want to discuss this further to ensure we don't accidentally foreclose options either by ``drying the well'' or perhaps simply in the way the missions are presented.

{\bf What costs are associated with these space-based GW missions?}

For the Moon-based missions, the interest of nations to develop and exercise their ability to perform work on the Moon is a powerful incentive to invest in science there. This brings a level of funding and government support that one is less likely to see for free-flyer missions unless they are also developing key technologies useful in space (with atom interferometry as an example). The cost of a Moon mission would likely not be burdened by the cost of development of capability of astronauts to play their roles, and the delivery of materials to the Moon may be part of a larger endeavor to gear up activities on the Moon. A sample initial mission to the Moon might cost of the order of 100 million dollars or euros.

The Moon missions under discussion offer the potential for geophysics and a better understanding of the Moon's composition. With Soundcheck, for instance, one can  study the distribution of moonquakes to much smaller magnitudes and reasonably mitigate the risk by extrapolating to a stationary background. In the process of building a good GW detector on the Moon, we will thus build a great tool to study the Moon itself. So, even in the worst-case scenario for the GW community where either the background is too noisy or GWs are not observed with the instruments as realized, one will still build tools that have scientific value. This sort of multi-use element can help spread interest, and thus funding, over a broader base, and also can compensate should the GW aspect of the mission not meet expectations. 

Some discussion indicated that the Moon is probably the only interesting target for a planet(oid)-based GW detector; of course high-sensitivity seismology of other planet(oid)s can also be of interest and could use techniques developed for LGWA or LILA.

The discussion brought to light that any multi-satellite free flyer would be a large mission for any of the space agencies, putting it in series with, e.g., LISA and with other intervening science-focused missions. This of course pushes out any likely timeline by order of a decade or more. The costs would likely all be counted against the free-flyer mission, including the $\sim 10\%$ launch cost. 

To note is that if there is a gap between a Pathfinder mission and the actual mission, some of the knowledge acquired with the Pathfinder becomes lost and must be re-learned. This is the case for LISA.

There was some discussion of the Chinese missions, and in particular TianQin, which would have a baseline (given the Earth orbits) that would enable good deci-Hz sensitivity. It appears that both TianQin and Taiji (which resembles closely LISA in science goals and configuration) have recently received significant funding for development, with a demonstrator planned for the coming year or so.

More generally, the deci-Hz community should continue its interaction with GWIC, which should represent a broad range of efforts in GW observation. One or more of the proposed Missions may be appropriate as members of GWIC; we note that TianQin joined in 2025.

\clearpage

\section{Wednesday morning --
Multi-messenger and multi-band astronomy in the deci-Hz band (discussion panel)}

\noindent
{\em Panel: Daniel D’Orazio, Chiara Mingarelli, Cole Miller, Jake Slutsky, Alessandra Corsi (Panel coordinator)}

\subsection{White dwarf mergers, double-degenerate model of type~Ia supernovae}
\label{sec:WDmergers}

Type Ia supernovae (SNe Ia), which are characterized observationally by their lack of hydrogen lines and (usually) the presence of strong silicon lines at peak flux, have for three decades been pillars of cosmography because their luminosity can be calibrated by their observed decay time~\cite{1993ApJ...413L.105P}.  They have also been understood for decades to be powered by thermonuclear explosions of WDs in binary systems, but it is still not certain whether the explosion is driven by accretion from a normal or giant companion (the single-degenerate scenario) or occurs due to the GW-driven coalescence of a binary WD system (the double-degenerate scenario; see~\cite{2025A&ARv..33....1R} for a recent summary). 

The double-degenerate scenario can be tested with unique power with deci-Hz GWs, and it also provides a prospect for a new type of multimessenger signal.   For the GW signal, we note that the maximum orbital frequency for WDs of typical masses $\sim 0.6-1~M_\odot$ is $\sim{\rm few}\times 10^{-2}$~Hz, i.e., close to the deci-Hz band.  At an expected deci-Hz double WD horizon of $\sim 50$~Mpc we expect several SNe~Ia per year (see~\cite{2025ApJ...992...16P, 2025arXiv250907849B} and Section~\ref{sec:DWD}).  Because near the end of their coalescence the WD separation would be comparable to their size, not only would gravitational radiation in the deci-Hz provide critical information such as the chirp mass, but tidal effects combined with WD stellar modeling will yield radii and possibly individual masses~\cite{2024MNRAS.531.4681Y}.

Double WD mergers also provide exciting opportunities for GW-EM multimessenger science. There are different perspectives about the expected time between a merger and the resulting SNe Ia (if, indeed, a SN Ia results; see~\cite{2025A&ARv..33....1R}).  For example, the authors of Ref.~\cite{2010ApJ...722L.157V} argues that a thick post-merger disk will accrete and provide enough compressional heating to initiate runaway fusion and a SN Ia within several hours, whereas the authors of Ref.~\cite{2018SCPMA..6149502S} propose a delay that is often millions of years because they suggest that angular momentum needs to be radiated away for runaway fusion to progress.  Resolution of this disagreement could provide invaluable information to SNe Ia modelers and might tighten constraints enough to improve cosmography with SNe Ia.  Regardless of the time to supernova, the merger itself will release $\sim 10^{50}$~erg and will be easily visible to the $\sim 30$~Mpc horizon of deci-Hz detectors.  This would provide a different type of multimessenger source even if no supernova is seen.

\subsection{Pre-localization of binary neutron star events, followed by kilonova and other electromagnetic emission}

A primary scientific motivation for deci-Hz observatories is the pre-localization of binary NS coalescences months or years prior to their merger in the frequency band of terrestrial detectors. While third-generation ground-based detectors like CE and ET will detect binary NS mergers with high signal-to-noise ratio, they typically observe only the final minutes of the inspiral. Conversely, deci-Hz instruments can track these systems for years, providing early warning alerts~\cite{Ajith:2024mie, Yelikar:2025jwh}.

This capability provides two distinct advantages. First, it enables the preparation of EM facilities. By refining the sky localization over a long baseline, deci-Hz detectors can alert EM observatories to the specific location and time of a merger, facilitating deep pre-merger imaging and the identification of potential precursors. Second, it allows for the optimization of observatory schedules. Knowledge of an upcoming ``golden'' event allows terrestrial observatories to adjust commissioning and maintenance schedules, ensuring maximum sensitivity at the moment of merger. Likewise, flagship class space-based observatories can be readied to immediately slew from nearby in the sky upon localization confirmations that should themselves be rapid, eliminating time-consuming manual target of opportunity and director's discretionary time processes that are long on the timescale of the transient involved.

The anticipated detection rates depend heavily on instrument design. Lunar concepts such as LGWA are projected to observe rates comparable to current LVK catalogs—approximately one to a few events per year within 50 Mpc~\cite{Harms_ESA}—while space-based interferometers like DECIGO could detect significantly more~\cite{Kawamura:2020pcg}. However, even a limited sample of binaries observed over a decade-long baseline would be scientifically valuable. Long duration observations allow for the measurement of orbital eccentricity and harmonics related to spin. While constraints on the nuclear EOS will primarily be derived from the tidal disruption phase observed by terrestrial detectors, the deci-Hz band provides critical initial conditions.

Crucially, deci-Hz observations break parameter degeneracies that limit ground-based analysis. Recent studies suggest that next-generation ground-based networks alone may only measure the symmetric mass ratio $\eta$ with high precision ($\delta \eta < 10^{-5}$) for approximately 100 binaries~\cite{Gupta:2023lga}. The addition of deci-Hz data allows for the precise determination of individual component masses. This is essential for distinguishing binary NS systems from NS-BH or binary BH systems prior to merger. Given that kilonovae and other EM counterparts are expected primarily from binary NS and specific NS-BH systems, the ability to strictly identify the system type and mass ratio in advance enables the efficient allocation of EM resources~\cite{Ronchini:2022gwk} and well as the potential detection of kilonovae from some of the large majority of mergers for which the Earth would not receive beamed gamma-ray emission (estimated at $\sim97\%$).

\subsection{Intermediate-mass black hole and extreme mass-ratio inspiral mergers in gas: tidal disruption events}

WDs can be tidally disrupted by IMBHs with masses from $\sim10\,M_\odot$ up to $\sim10^5\,M_\odot$~\cite{Maguire:2020lad}.
Recently, a candidate has been identified~\cite{Li:2025mae}.
Tidal disruption of WDs produces multimessenger signals, spanning a range from GWs to EM waves in the X-ray, optical, and ultraviolet spectra. Since WDs are composed of degenerate matter, they may also cause a nuclear detonation.
On the other hand, WDs are swallowed whole by supermassive BHs without associated EM counterparts due to the smaller tidal radius relative to the BH horizon for BH masses $\gtrsim10^5\,M_\odot$.
The rate of WD disruption by IMBHs depends on astrophysical models, ranging from $\sim0.01$--$100\,\rm yr^{-1}$, and these events can be detectable out to $\sim200\,\rm Mpc$~\cite{Sesana:2008zc}.
The orbital properties of the hyperbolic orbit place these GW sources in the deci-Hz window and make them essential sources for deci-Hz GW observatories~\cite{Toscani:2025uar}.

\subsection{Multi-band: formation channels for binary black holes}

Stellar-mass compact object binaries that end their lives in the LVK bands will pass through deci-Hz frequencies on their way to merger (see e.g.~\cite{Sesana:2016ljz}), presenting unique opportunities for probing astrophysical environments and formation scenarios. This is because at these pre-merger, deci-Hz frequencies:
\begin{enumerate}
    \item Environmental effects can have a larger imprint than they would near merger,
    \item Binaries can retain higher eccentricity, 
    \item Dynamical capture events can form with peak GW power emitting in band.
\end{enumerate}

The first two points follow simply from the steep increase in GW emission with decreasing semi-major axis, causing GW emission to dominate and eccentricity to damp. The first point is also helped by the longer time the binary spends at deci-Hz frequencies, allowing, e.g., line of sight acceleration or tidal effects from a companion to imprint on the waveform \citep{Samsing:2024syt}. 
These arguments are of course even stronger for stellar-mass binaries in the LISA band, however, a higher detectability rate could benefit the deci-Hz in this regard \citep{Gerosa:2019dbe,Iacovelli:2025kwn}, in addition to sampling another point in the binary evolution.

A longer time in band also enables a number of other multi-band opportunities 
for detecting environmental waveform modulations
in the deci-Hz. For example, Stegmann {\it et al.}~\cite{StegmannZwick+2024} show that modulations of GWs from stellar mass binary BH inspirals could be used to detect inspiralling $\mathcal{O}(10^7-10^8)M_{\odot}$ SMBHBs in the binary BH's host galaxy. In another example, strongly lensed binary BH inspirals could encode their center-of-mass proper motions through waveform modulations that are caused by the different Doppler shifts of each lensed ``image''~\cite{SamsingDan_LensPM+2024}.

The third point arises because highly eccentric systems emit peak GW power at a frequency far above that of a circular binary. Binaries formed especially through single-single, bremsstrahlung-type capture in globular clusters will form with an eccentricity high enough for the peak frequency to first appear high of the LISA band, within the deci-Hz band (see, e.g., Fig.~5 of~\cite{Samsing:2019dtb} and earlier work on the formation of such capture events in nuclear star clusters \citep{KocsisLevin:2012, OLearyKocsisLoeb:2009}.) 
Because highly eccentric orbits evolve by circularizing at a nearly constant pericenter distance, and because the peak frequency is approximately the orbital frequency of a circular orbit at pericenter, such systems will emit peak power near the formation peak frequency until they circularize. 
The time for these binaries to circularize and merge can be of order a year for the highest frequency systems (e.g., $f=0.1$Hz, $M=100\,M_\odot$, $e_0=0.999$), so deci-Hz observations could see the birth of such bremsstrahlung capture events. For most of the longer-lived, lower frequency systems, a deci-Hz detector could at least observe a highly eccentric population indicative of formation with peak frequency above the LISA band.
Furthermore, high signal-to-noise ratio events could conceivably be simultaneously detectable in both deci-Hz and 3G detectors. 
Hence, the deci-Hz in a multi-band setting with LISA and ground-based detectors could offer a powerful discriminator of dynamical formation pathways.

\subsection{Probes of primordial gravitational wave backgrounds across frequency bands}

Astrophysical effects typically become more important at low GW frequencies. For example, environmental coupling can drive stronger dephasing at low frequencies than at high frequencies, complicating attempts to interpret the signal as purely GW driven.

A central question is whether astrophysical and cosmological anisotropies have distinct physical origins, and whether those origins can be disentangled observationally. One approach is foreground mitigation: either (i) individually resolve and subtract astrophysical sources when possible, or (ii) use population-level information to model and subtract the astrophysical foreground statistically.

An important nuance is the role of resolved GW sources. Individually resolved events can provide direct information about binary parameters and the relevant astrophysical systematics, potentially improving subtraction and inference. However, resolved sources are not strictly required if the correlated signal and the foreground exhibit sufficiently distinct statistical signatures, as demonstrated by existing population-based analyses in PTA studies.

\clearpage

\section{Conclusions}
\label{sec:conclusions}

This workshop highlighted that GW sources in the deci-Hz band are plentiful and that their observation is an important experimental target.

Further progress on the experimental side will require a White Paper-type analysis that compares projected sensitivities of alternative or complementary proposals for the deci-Hz band. Such a study is currently under way for free flyers in Europe (GW-Space~2050, see Section~\ref{sec:GWSpace2050}) and focuses on the post-LISA environment. If we take one lesson from this workshop, it is that  we need a parallel ``GW-Moon~2050'' study  to clarify and consolidate future lunar or cislunar projects, in parallel to the current  free flyer studies of GW-Space~2050. 

The lunar horizon is on a shorter time-scale than the 2065 call anticipated in Voyage~2050 for a post-LISA option at the proposed cadence of ESA-L missions. Indeed, lunar planning by NASA and ESA  is happening now. The first step announced in January 2026 is a seismometer project for ARTEMIS-4 to study lunar seismological backgrounds. Our community must be ready for the next phase with strong ESA and NASA interest. A GW-Moon~2050 study is needed to set out an optimal way forward.  

\clearpage

\section*{Acknowledgments}

E.~Berti, F.~Iacovelli, and K.~Kritos are supported by NSF Grants No.~AST-2307146, No.~PHY-2513337, No.~PHY-090003, and No.~PHY-20043, by NASA Grant No.~21-ATP21-0010, by John Templeton Foundation Grant No.~62840, by the Simons Foundation [MPS-SIP-00001698, E.Berti], by the Simons Foundation International [SFI-MPS-BH-00012593-02], and by Italian Ministry of Foreign Affairs and International Cooperation Grant No.~PGR01167. 
A.Buonanno's research is supported in part by the European Research Council (ERC) Horizon Synergy Grant “Making Sense of the Unexpected in the Gravitational-Wave Sky” grant agreement no. GWSky–101167314.
J.~Harms receives financial support from the Italian Space Agency (ASI) under Grant No. 2025-29-HH.0.
F.~Iacovelli is supported by a Miller Postdoctoral Fellowship.
M.\ Kamionkowski was supported by NSF Grant No.\ 2412361, NASA ATP Grant No.\ 80NSSC24K1226, and  the Templeton Foundation.
K.~Komori was supported by JSPS KAKENHI Grant Number JP22K14066.  
K.~Kritos is further supported by the Onassis Foundadtion - Scholarship ID: F ZT 041-1/2023-2024.
A.Maselli acknowledges support from the ITA-USA Science and Technology Cooperation program, supported by the Ministry of Foreign Affairs of Italy (MAECI), and from MUR PRIN Grants No.~2022-Z9X4XS and No.~2020KB33TP.
C.~M.~F.~Mingarelli\ was supported in part by the National Science Foundation under Grants No.~NSF PHY-1748958, AST-2106552, and NASA LPS 80NSSC24K0440. 
B.~S.~Sathyaprakash is supported by NSF Grants No.~AST-2307147, No.~PHY-2308886 and No.~PHY-2309064. 
D.~H.~Shoemaker\ was supported by NASA award 80NSSC23K1242, and NSF Grants PHY-2309064 and PHY-18671764464.

\clearpage

\bibliographystyle{apsrev4-1}
\bibliography{references}

\end{document}